\documentclass[aps,prx,reprint,superscriptaddress,longbibliography]{revtex4-2}
\usepackage[T1]{fontenc}
\usepackage{lmodern}
\usepackage[version=4]{mhchem}
\usepackage{amsmath,amssymb,bm,graphicx}
\usepackage[colorlinks=true,linkcolor=blue,citecolor=blue,urlcolor=blue]{hyperref}
\hypersetup{pdftitle={Hidden Rotational Dynamics from Crystal Geometry},pdfauthor={Rule Yi, Vincent P. Flynn, Benedetta Flebus}}
\graphicspath{{figures/}{./}}
\newcommand{\bu}{\hat{\mathbf u}}
\newcommand{\bp}{\hat{\mathbf p}}

\usepackage{xcolor}
\definecolor{darkgreen}{rgb}{0.0, 0.5, 0.0}

\begin{document}
\title{Hidden Rotational Dynamics from Crystal Geometry}
\author{Rule Yi}
\thanks{These authors contributed equally to this work.}
\affiliation{Department of Physics, Boston College, Chestnut Hill, Massachusetts 02467, USA}

\author{Vincent P. Flynn}
\thanks{These authors contributed equally to this work.}
\affiliation{Department of Physics, Boston College, Chestnut Hill, Massachusetts 02467, USA}
\affiliation{Department of Physics and Astronomy, Dartmouth College, Hanover, New Hampshire 03755, USA}

\author{Benedetta Flebus}
\affiliation{Department of Physics, Boston College, Chestnut Hill, Massachusetts 02467, USA}
\date{\today}

\begin{abstract}
 From hidden spin order to electronic loop currents and frustrated fluctuations, unconventional quantum matter reveals the rich patterns that electronic and magnetic degrees of freedom form across space and time. Recent advances in chiral phononics have extended this perspective to lattice vibrations, raising the question of whether the lattice itself can generate spatially organized rotational correlations. Here we show that crystal geometry provides an autonomous mechanism for generating equilibrium angular-momentum circulation and staggered antisymmetric displacement interactions, independently of coupling to electronic or magnetic order. Our framework applies broadly to crystalline systems hosting locally symmetry-lowered environments embedded within a higher-symmetry global structure. We identify how directional elastic forces organize displacement correlations and how eliminating internal coordinates generates staggered antisymmetric interactions—a phononic analogue of Dzyaloshinskii–Moriya exchange—even in globally centrosymmetric crystals. By separating intersite angular-momentum transfer from crystalline restoring torques, we show that both elastic anisotropy and the induced antisymmetric interactions together 
drive oppositely circulating angular-momentum currents on neighboring plaquettes, despite vanishing mean angular momentum at every site. Sustained by quantum and thermal fluctuations, this circulation preserves time-reversal and inversion symmetry and requires neither external driving nor spontaneous symmetry breaking. Our results establish lattice geometry as a microscopic source of spatially organized angular-momentum dynamics. 
\end{abstract}
\maketitle

\section{Introduction}
\label{sec:Intro}
Recent theoretical and experimental advances have established phonons as carriers of angular momentum~\cite{zhang2014angular,zhang2015chiral,zhu2018observation,tauchert2022polarized,ueda2023chiral,zhang2025measurement,juraschek2025chiral}, uncovering the long-overlooked role of lattice rotations in shaping the properties of quantum materials~\cite{juraschek2017dynamical,basini2024terahertz}. Concurrently, several works have shown how lattice vibrations can acquire angular momentum through coupling to electronic and magnetic degrees of freedom, revealing that spins can transfer angular momentum to the lattice~\cite{zhang2014angular,tauchert2022polarized}, while magnetic order can split phonon modes of opposite circular polarization~\cite{Bonini2023ChiralPhonons} and control the handedness of lattice motion~\cite{Wu2025MagneticSwitching}. Coupling to electrons has likewise been predicted to generate circularly polarized optical modes through molecular Berry curvature~\cite{Saparov2022MolecularBerry}, imprint orbital textures on phonons~\cite{ChatzichrysafisMook2026}, and generate  phonon angular momentum under driving~\cite{Yokoyama2025PhononEdelstein}. 

In turn, rotational lattice dynamics can influence electronic and magnetic degrees of freedom, enabling phenomena ranging from magnetization switching~\cite{Davies2024} to the generation of spin and orbital currents~\cite{Kim2023CPASS,Nabei2026OrbitalSeebeck}. Experiments have also revealed angular-momentum transfer between coherently excited lattice modes~\cite{Minakova2026Transfer} and field-induced phonon thermal Hall responses in nonmagnetic crystals~\cite{Jin2025UniversalThermalHall}.
These developments have motivated approaches that treat lattice motion and its rotational character more explicitly. Recent first-principles theories place spin and lattice dynamics on equal footing~\cite{Ren2024Adiabatic}, while generalized continuum theories extend the description of crystalline dynamics to account for phonon angular momentum~\cite{Matsuo2026Continuum}. \begin{figure}[t]
    \centering
    \includegraphics[width=\linewidth]{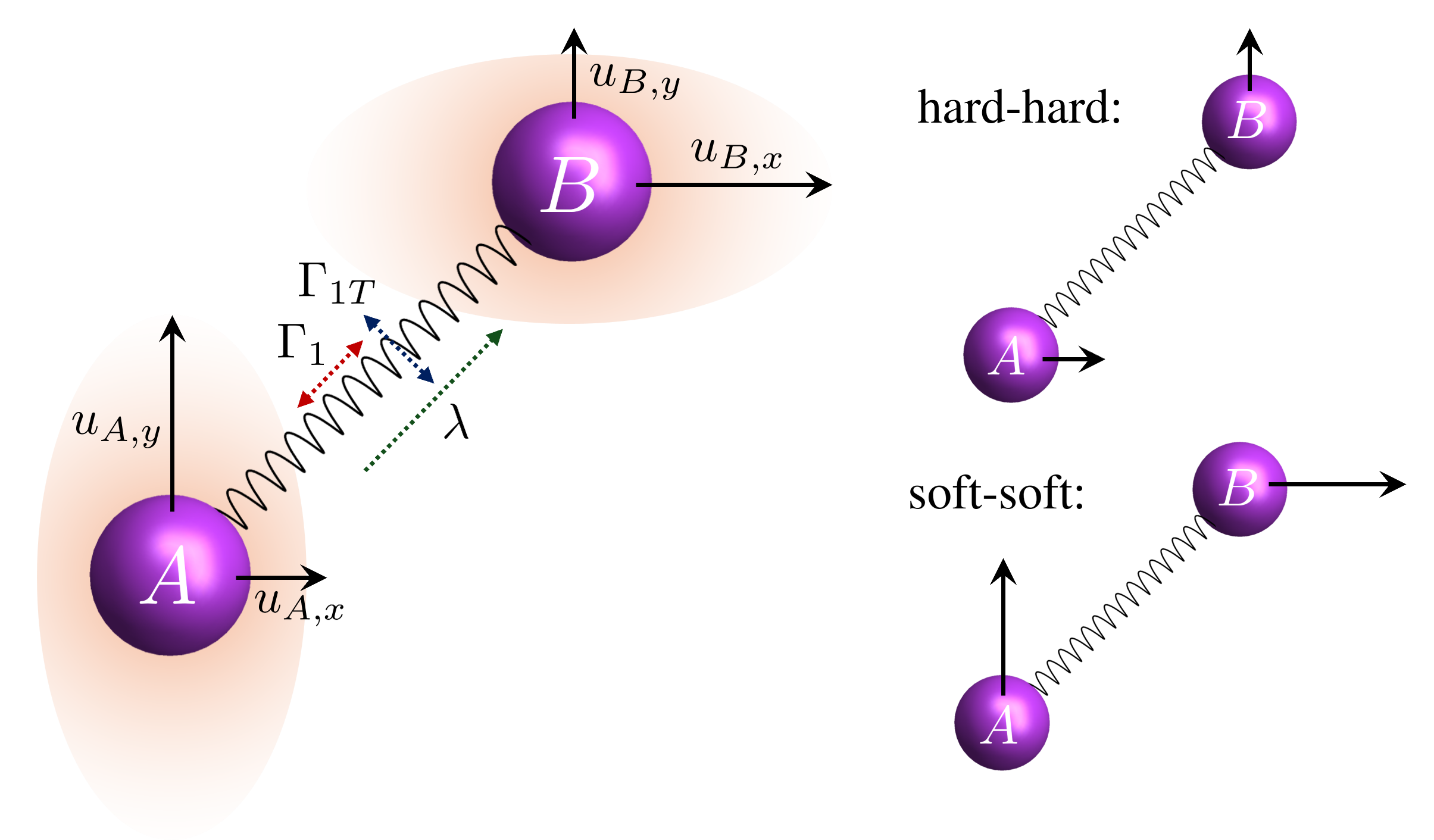}
    \caption{Schematic illustration of equilibrium angular momentum exchange. The local elastic environments at sites $A$ and $B$ define perpendicular soft displacement directions, associated with $ u_{Ay}$ and $ u_{Bx}$, and complementary hard directions, associated with $ u_{Ax}$ and $ u_{By}$. A diagonal anisotropic spring couples the hard--hard and soft--soft displacement pairs with the same coefficient, proportional to $\Gamma_1-\Gamma_{1T}$. The softer pair develops a larger displacement correlation, yielding $\langle \hat u_{Ay}\hat u_{Bx}\rangle \neq \langle \hat u_{Ax}\hat u_{By}\rangle$. Their opposite contributions to angular momentum exchange therefore
do not cancel, allowing a finite mean current $\langle \hat J_{A\to B}\rangle$ even when the mean local angular
momentum vanishes at each site. In the isolated pair, local restoring torques balance this exchange. The label $\lambda$ denotes the additional antisymmetric displacement coupling in Eq.~\eqref{eq:H}, whose contribution to the current is discussed in detail in Sec.~\ref{sec:EC}.}
    \label{toymodel}
\end{figure} 
Growing recognition of the role of lattice rotations has also led to the proposal of an equilibrium phase whose order parameter is the phonon angular momentum itself, with spin fluctuations mediating the interactions that drive its spontaneous ordering~\cite{Fahrensohn2026Phonomagnet}. Yet nontrivial rotational dynamics require neither spin coupling nor spontaneous order. As shown by recent work~\cite{YiWilliamsFlebus2026}, even crystals preserving inversion and time reversal can exhibit equilibrium phonon angular-momentum fluctuations both in the ground state and at finite temperature. These fluctuations are sourced by the directional character of crystalline elasticity, which generally lacks the internal $U(1)$ symmetry generated by phonon angular momentum. Because longitudinal and transverse bond deformations experience different restoring stiffnesses, rotating the atomic displacements together relative to the fixed lattice changes the elastic energy, allowing phonon angular momentum to fluctuate dynamically even when inversion and time reversal constrain its equilibrium mean to vanish~\cite{Coh2023Classification}. The ubiquity of these fluctuations naturally invites a search for hidden rotational structure in the spatial correlations of lattice motion, much as spin correlations reveal collective behavior not captured by the mean spin configuration~\cite{PencLauchli2011,FernandesOrthSchmalian2019,AndreevGrishchuk1984,ChaikinLubensky1995,Henley2010,Balents2010}.

Here we show that such structure can take the form of equilibrium angular-momentum circulation. Specifically, we find that crystal geometry organizes quantum and thermal displacement correlations into a finite mean exchange of angular momentum between sites, with crystal symmetry fixing the circulation pattern and elastic couplings determining its sense.
\begin{figure*}[t]
\includegraphics[width=1\textwidth]{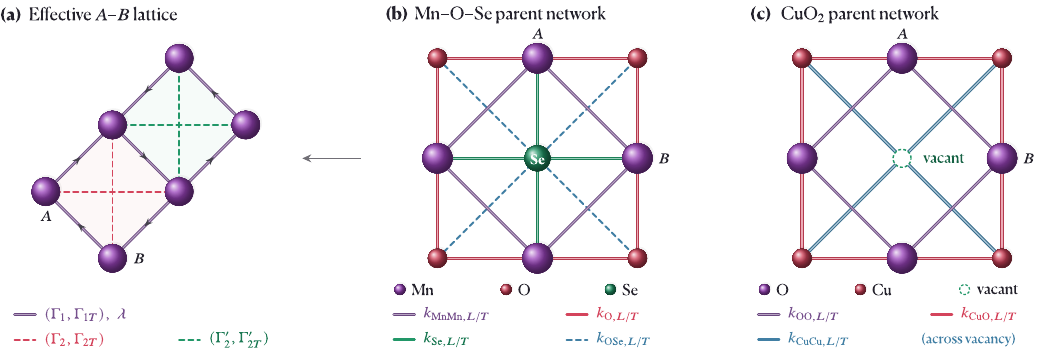}
\caption{Effective lattice and representative parent elastic networks.
(a) Effective $A$--$B$ lattice capturing the rotation-related local
environments in both (b) and (c). The $A$ and $B$ sites correspond
to Mn in (b) and O in (c). Diagonal NN bonds carry
$(\Gamma_1,\Gamma_{1T})$ and $\lambda$; the two axial NNN families
carry $(\Gamma_2,\Gamma_{2T})$ and $(\Gamma_2',\Gamma_{2T}')$.
The arrows indicate opposite senses of angular momentum circulation
around the two inequivalent plaquette families. 
(b) A planar network inspired by
$\mathrm{La}_2\mathrm{O}_3\mathrm{Mn}_2\mathrm{Se}_2$ contains
Mn--Mn, Mn--O, Mn--Se, and O--Se elastic interactions, each
described by longitudinal and transverse stiffnesses. The Se site
lies at the cell center. Relaxing the O and Se coordinates at
fixed Mn displacements yields the effective elastic couplings
represented in (a).
(c) A \ce{CuO2} parent network illustrates the analogous
rotation-related environments of the two oxygen sublattices,
with Cu-centered and vacant plaquettes providing the two
inequivalent local environments. }
\label{fig:model}
\end{figure*}
Our results apply broadly to harmonic lattices in which locally anisotropic environments are embedded within a higher-symmetry crystal and do not rely on the presence of electronic or magnetic order. The origin of this equilibrium angular-momentum exchange can be understood through the simple mechanical picture illustrated in Fig.~\ref{toymodel}. Consider two particles, \(A\) and \(B\), with displacements \(\mathbf{u}_A\) and \(\mathbf{u}_B\) from equilibrium. They are connected by a diagonal spring and confined more weakly along \(y\) and \(x\), respectively. Although the spring couples \(u_{Ay}\) to \(u_{Bx}\) and \(u_{Ax}\) to \(u_{By}\) with equal strength, the different restoring forces give rise to correlations of unequal magnitude: \(\langle u_{Ay}u_{Bx}\rangle\) is larger in magnitude than \(\langle u_{Ax}u_{By}\rangle\). These correlations enter the average cross product of neighboring displacements with opposite signs and therefore do not cancel, producing a finite angular-momentum current through the angular-momentum-conserving part of the bond interaction.

The anisotropic interactions that shape these correlations also generate crystalline restoring torques: their elastic energy depends on the orientation of the displacements relative to the crystal axes and therefore lacks internal \(U(1)\) symmetry. In the two-particle model of Fig.~\ref{toymodel}, equilibrium requires the total mean restoring torque at each site to balance the exchange current. In a crystal, geometry can organize the bond currents into closed loops, with incoming and outgoing currents balancing at every site. Because angular-momentum exchange is even under time reversal, such circulation can coexist with vanishing mean angular momentum at every site. Zero-point motion supplies the underlying fluctuations in the phonon vacuum, and the same mechanism persists in thermal equilibrium.

As a concrete illustration of how these correlations shape the rotational dynamics of a crystal, we consider an inverse Lieb lattice with two sublattices related by a \(90^\circ\) crystal rotation and inequivalent plaquette environments [Fig.~\ref{fig:model}].  This geometry captures the rotation-related local environments found at the Mn sites of the candidate altermagnet La$_2$O$_3$Mn$_2$Se$_2$ and at oxygen sites of CuO$_2$ planes. Here, we find that symmetric, direction-dependent elastic interactions alone give rise to bond currents that form oppositely circulating loops on neighboring plaquettes, while preserving inversion, time reversal, and fourfold rotational symmetry.

The implications of this geometry, however, extend beyond circulation to the structure of effective lattice Hamiltonians. Our microscopic analysis further shows that eliminating internal ionic coordinates from a parent elastic network generates staggered Dzyaloshinskii--Moriya-like interactions between the retained displacements, even when every microscopic bond tensor is symmetric. Antisymmetric displacement interactions have previously been explored in connection with structural distortions and electronic contributions to interatomic forces~\cite{Zhao2021,Chen2022}. Our results reveal a geometric route to these interactions within a purely harmonic lattice, with their staggered pattern enabled by inequivalent local bonding environments that lack bond-centered inversion while preserving global inversion.  This antisymmetric interaction also provides a second contribution to angular-momentum circulation: a displacement on one site induces a perpendicular force on its neighbor, allowing correlations in their relative alignment to contribute directly to angular-momentum transfer. Unlike symmetric anisotropy, the DMI-like term is itself invariant under internal rotations. In a circular basis, it introduces phases into the angular-momentum-conserving exchange, whose accumulation around closed paths is encoded in \(U(1)\) Wilson loops~\cite{Wilson1974}. These loop phases alternate between neighboring plaquettes and reverse between the two circular-polarization sectors, connecting this contribution to geometric descriptions of equilibrium spin currents~\cite{SchuetzKollarKopietz2003,FlynnFlebus2026}. Circulation sustained by symmetric anisotropy, by contrast, can persist even when these loop phases are trivial.

Taken together, our results show that crystal geometry can generate staggered antisymmetric displacement interactions and equilibrium angular-momentum circulation while preserving inversion and time-reversal symmetry. The spatial pattern of this circulation recalls the electronic loop-current phases proposed for cuprates and kagome metals, although their charge currents break time reversal~\cite{Varma2006,Feng2021ChiralFlux}. A closer symmetry analogue is provided by proposed staggered spin-current states, in which spin-up and spin-down electrons countercirculate while preserving time reversal~\cite{Nersesyan1991,Nayak2000,CapponiAssaad2007,Hsu2011}. Whereas these currents transport the intrinsic spin of electrons, the circulation found here arises from correlated ionic motion in a lattice whose inversion and time-reversal symmetries constrain the angular momentum of every nondegenerate phonon mode to vanish~\cite{Coh2023Classification}. Rotational lattice correlations can thus precede electronic and magnetic ordering, carrying a definite spatial and polarization structure in the symmetric phase. Through coupling to electrons and spins, this structure could shape the effective interactions governing competing instabilities, suggesting that lattice rotations may help select the order that ultimately emerges.

 This work is organized as follows. In Sec.~\ref{sec:Model}, we introduce the lattice model and derive its elastic couplings from a microscopic parent network. In Sec.~\ref{sec:AM}, we separate angular momentum-conserving exchange from the anisotropic interactions responsible for crystalline restoring torques. In Sec.~\ref{sec:Current}, we define the bond-current operator and establish the symmetry constraints on the currents and their local balance with restoring torques. In Sec.~\ref{sec:EC}, we evaluate the equilibrium current from quantum and thermal displacement correlations and identify the two routes to circulation provided by elastic anisotropy and antisymmetric exchange. In Sec.~\ref{sec:WL}, we construct the exchange Wilson loops and explain why anisotropic correlations can sustain circulation even when these loops are trivial. Finally, Sec.~\ref{sec:discussion} summarizes our findings and discusses their implications and experimental prospects.

\section{Model}
\label{sec:Model}
 Our analysis applies broadly to crystals hosting environments of reduced local symmetry within a higher-symmetry global structure, where the spatial arrangement of local anisotropies can organize angular momentum exchange without breaking the crystal symmetries. These environments may be encoded directly in the microscopic bonding geometry or become apparent only upon resolving lattice motion into collective and internal degrees of freedom. 
 To make this connection concrete, we study a minimal two-dimensional model preserving inversion, time reversal, and fourfold rotation. Its geometry captures local bonding motifs and the symmetry of in-plane vibrations in tetragonal materials, including cuprates that host high-temperature superconductivity, electronic nematicity, and charge-density-wave correlations~\cite{Tabis2014ChargeOrder,Murayama2019DiagonalNematicity}.
Specifically, we consider a planar harmonic lattice with plane group $p4mm$ and time-reversal symmetry. Its Bravais vectors are $a\hat{\mathbf{x}}$ and $a\hat{\mathbf{y}}$, and its two dynamical sites have basis positions 
$
\boldsymbol{\delta}_A = (a/2,0),
$ and 
$\boldsymbol{\delta}_B = (0,a/2)$, where the labels $A$ and $B$ denote symmetry-related sublattices of the same atomic species. Inequivalent local environments occupy the plaquette centers at $(0,0)$ and $(a/2,a/2)$, as illustrated in Fig.~\ref{fig:model}(a).

The 
crystal symmetries are generated by crystal translations, the fourfold rotation $C_4$, and the mirror $m_x$:
\begin{equation*}
C_4 : (x,y) \mapsto (-y,x),
\qquad
m_x : (x,y) \mapsto (-x,y).
\end{equation*}
These operations preserve the two families of plaquette environments separately. The lattice therefore admits global in-plane inversion $\mathcal{P}=C_4^2$, which maps $(x,y)$ to $(-x,-y)$. Inversion about a nearest-neighbor $AB$-bond midpoint, such as $(a/4,a/4)$, instead exchanges the inequivalent environments and is thus not a symmetry.

The \ce{Mn2OSe2} layers of the layered oxyselenide candidate altermagnet $\mathrm{La}_2\mathrm{O}_3\mathrm{Mn}_2\mathrm{Se}_2$ provide a structural reference. As shown in Fig.~\ref{fig:model}(b), the local environments of
the Mn sites are orthorhombic, with principal axes interchanged
by a fourfold rotation~\cite{Ni2010,Wei2025,GarciaGassull2026,Asai2026}. An analogous local arrangement occurs, e.g.,  for the two planar oxygen sublattices, $\mathrm{O}_x$ and
$\mathrm{O}_y$, in the $\mathrm{CuO}_2$ planes of tetragonal
cuprates~\cite{Emery1987,Saunderson2025Oxygen},  as shown in  Fig.~\ref{fig:model}(c). These examples motivate the geometry of the model.

A site is labeled by $i=(\mathbf R,s)$, where $\mathbf R$ is
a Bravais lattice vector and $s$ labels a site within the unit
cell. Its equilibrium position is
$\mathbf R_i=\mathbf R+\boldsymbol\delta_s$.
We denote its position operator by $\hat{\mathbf r}_i$ and
its physical momentum by $\hat{\mathbf P}_i$, and define the
displacement from equilibrium as
$\delta\hat{\mathbf r}_i=\hat{\mathbf r}_i-\mathbf R_i$.
For a site of mass $m_i$, we introduce the mass-weighted
canonical variables
\begin{equation}
\hat{\mathbf u}_i=\sqrt{m_i}\,\delta\hat{\mathbf r}_i,
\qquad
\hat{\mathbf p}_i=\frac{\hat{\mathbf P}_i}{\sqrt{m_i}},
\end{equation}
which satisfy
$[\hat u_{i\alpha},\hat p_{j\beta}]
=i\hbar\delta_{ij}\delta_{\alpha\beta}$,
with $\alpha,\beta\in\{x,y\}$.
We retain terms quadratic in the displacements and restrict the interactions to nearest-neighbor (NN) bonds and axial next-nearest-neighbor (NNN) bonds, which run along the $x$ and $y$ directions. 
Imposing invariance under uniform translations $\mathbf{\hat u}_i\mapsto \mathbf{\hat u}_i+\mathbf{a}$ and enforcing crystal symmetries,  we write the Hamiltonian as
\begin{align}
\hat H={}&
\frac12\sum_i\bp_i^2
+\frac12
\sum_{\langle ij\rangle
}
(\bu_j-\bu_i)^T 
K_{ij}(\bu_j-\bu_i)
\notag\\
&-\lambda\sum_{A\to B}\eta_{ij}(\bu_i\times\bu_j)_z.
\label{eq:H}
\end{align}
The elastic response of the bond connecting sites $i$ and $j$ is
described by the symmetric tensor
\begin{equation}
K_{ij}
=
\Gamma_{ij,T}\mathbb I_2
+
(\Gamma_{ij}-\Gamma_{ij,T})
\mathbf e_{ij}\mathbf e_{ij}^{T},
\label{eq:elastic-tensor}
\end{equation}
where $\mathbb I_2$ is the two-dimensional identity matrix and
$\mathbf e_{ij}$ is the unit vector along the equilibrium bond.
The coefficients $\Gamma_{ij}$ and $\Gamma_{ij,T}$ are the
mass-weighted stiffnesses for relative displacements parallel and perpendicular
to the bond, respectively.
Each undirected bond $\langle ij\rangle$ is counted once in the elastic sums. There are three such families: (i) the diagonal $A$--$B$ bonds whose stiffnesses we denote by $(\Gamma_1,\Gamma_{1T})$; (ii) the horizontal $A$--$A$ and vertical $B$--$B$ bonds whose (shared) stiffnesses we denote by $(\Gamma_2,\Gamma_{2T})$; and (iii) the vertical $A$--$A$ and horizontal $B$--$B$ bonds whose (shared) stiffnesses we denote by $(\Gamma^\prime_2,\Gamma^\prime_{2T})$. 
The term proportional to $\lambda$ is an antisymmetric displacement interaction, i.e., the phonon analogue of the Dzyaloshinskii--Moriya interaction (DMI)~\cite{Dzyaloshinsky1958,Moriya1960}, with related antisymmetric displacement couplings studied as electric DMI~\cite{Chen2022,Zhao2021}. Importantly, this term is allowed by the broken inversion symmetry centered at an $A-B$ bond midpoint, i.e., the same symmetry breaking resulting from inequivalent sublattice bond environments. By contrast, inversion symmetry forbids such an interaction for axial NNN bonds. In this term, each NN bond is counted once
and oriented from $A$ to $B$.
For $i\in A$ and $j\in B$, we write
\begin{equation}\label{eq: bondvec}
\mathbf R_j-\mathbf R_i
=\frac{a}{2}(\eta_x,\eta_y),
\qquad
\eta_x,\eta_y=\pm1,
\end{equation}
and define the 
sign factor
$\eta_{ij}=\eta_x\eta_y$. 
Here, the four sign factors associated with the bonds incident
on an $A$ site sum to zero, thus ensuring global translation invariance.

To connect these effective couplings to microscopic
elasticity, we introduce the parent network shown in Fig.~\ref{fig:model}(b).
Each unit cell contains the two Mn sites and internal O
and Se sites at
$\boldsymbol\delta_{\mathrm O}=(0,0)$ and
$\boldsymbol\delta_{\mathrm{Se}}=(a/2,a/2)$.
We write its Hamiltonian in terms of the physical
displacements $\delta\hat{\mathbf r}_i$ and momenta
$\hat{\mathbf P}_i$, allowing the atomic masses $m_i$
to depend on the species:  
\begin{equation}
\begin{aligned}
\hat{H}_{\mathrm{parent}}
&= \sum_i \frac{\hat{\mathbf{P}}_i^{\,2}}{2m_i}
+ \frac{1}{2}\sum_{\{ij\}}
(\delta\hat{\mathbf{r}}_j-\delta\hat{\mathbf{r}}_i)^T
K_{ij}^{\mathrm{parent}}
(\delta\hat{\mathbf{r}}_j-\delta\hat{\mathbf{r}}_i),
\\
K_{ij}^{\mathrm{parent}}
&= k_{ij,T}\mathbb I_2
+ (k_{ij,L}-k_{ij,T})\mathbf{e}_{ij}\mathbf{e}_{ij}^T,
\end{aligned}
\end{equation}
where $i$ runs over all Mn, O, and Se sites, and each
unordered bond $\{ij\}$ is counted once.
The coefficients $k_{ij,\mathrm L}$ and
$k_{ij,\mathrm T}$ are physical longitudinal and
transverse spring stiffnesses, with units of energy
per length squared.
The spring families include Mn--Mn, Mn--O, Mn--Se, and O--Se bonds with stiffnesses denoted by $k_{\mathrm {MnMn},L/T}$,
$k_{\mathrm O,L/T}$, $k_{\mathrm{Se},L/T}$, and
$k_{\mathrm{OSe},L/T}$, respectively, where $L$ and $T$
refer to longitudinal and transverse bond deformations. The Mn--O bonds lie along $x$ for $A$ and $y$ for $B$; the Mn--Se directions are interchanged.
We denote by $\Phi_{i\alpha,j\beta}$ the full real-space
force-constant matrix, defined as the Hessian of the
parent elastic energy with respect to the physical
displacements. The corresponding mass-weighted matrix
has elements
\begin{equation}
D^{\mathrm{parent}}_{i\alpha,j\beta}
=
\frac{\Phi_{i\alpha,j\beta}}{\sqrt{m_i m_j}}.
\end{equation}

For both the parent and
effective lattices, we use the Fourier convention
\begin{equation}
\hat{\mathbf u}_{\mathbf R s}
=
\frac{1}{\sqrt{N_c}}
\sum_{\mathbf k}
e^{i\mathbf k\cdot(\mathbf R+\boldsymbol\delta_s)}
\hat{\mathbf u}_s(\mathbf k),
\label{eq:fourier-convention}
\end{equation}
where $N_c$ is the number of unit cells.
We then minimize the elastic energy with respect to the O
and Se coordinates, holding the Mn displacements fixed, to obtain the static elastic response.
We partition $D^{\mathrm{parent}}(\mathbf k)$ into
a retained Mn sector, ordered as $(A_x,A_y,B_x,B_y)$,
and an internal sector, ordered as
$(\mathrm O_x,\mathrm O_y,\mathrm{Se}_x,\mathrm{Se}_y)$.
Denoting these sectors by $\mathrm{Mn}$ and
$\mathrm{int}$, respectively, gives
\begin{equation}
\begin{aligned}
D_{\mathrm{stat}}(\mathbf k)
=&
D_{\mathrm{Mn},\mathrm{Mn}}
\\
-&
D_{\mathrm{Mn},\mathrm{int}}
D_{\mathrm{int},\mathrm{int}}^{-1}
D_{\mathrm{int},\mathrm{Mn}},
\end{aligned}
\label{eq:static-reduction}
\end{equation}
where all blocks on the right-hand side belong to
$D^{\mathrm{parent}}(\mathbf k)$ and are evaluated at
the same wave vector.
This static reduction determines the relaxed elastic
stiffnesses~\cite{Guyan1965}. We retain
its NN and axial NNN components to construct
Eq.~\eqref{eq:H}, with onsite terms fixed by the acoustic sum rule. The reduction
generally also generates longer-range interactions, which are outside the
retained model. The effective antisymmetric interaction $\propto \lambda$ arises because propagation through
differently oriented anisotropic springs can produce a directed intersite
tensor with an antisymmetric Cartesian component. The O and Se environments
give unequal contributions to that component. For example, take
$k_{\mathrm O,T}=k_{\mathrm{Se},T}=0$ and
\[
\max\!\left(
|k_{\mathrm{OSe},L}|,|k_{\mathrm{OSe},T}|
\right)
\ll
\min\!\left(k_{\mathrm O,L},k_{\mathrm{Se},L}\right).
\]
To second order in the O--Se stiffnesses, the leading
antisymmetric interaction is
\begin{align}
\lambda \simeq
\frac{
\left(k_{\mathrm{OSe},L}^{2}-k_{\mathrm{OSe},T}^{2}\right)
\left(k_{\mathrm{O},L}-k_{\mathrm{Se},L}\right)
}{
32M k_{\mathrm{O},L}k_{\mathrm{Se},L}
},
\label{eq:lambda}
\end{align}
while the contrast between the two axial NNN bond families is
\begin{align}
\Gamma_2-\Gamma^\prime_2 \simeq
\frac{k_{\mathrm{O},L}-k_{\mathrm{Se},L}}{2M},
\label{eq:Gamma}
\end{align}
where $M$ is the mass of a Mn atom. The more general leading expressions are given in Appendix~\ref{app:parent}.

These expressions reveal how the same microscopic environment contrast enters both circulation mechanisms. Unequal Mn--O and Mn--Se stiffnesses distinguish the two axial bond families and, in this limit, establish perpendicular hard and soft directions on the two Mn sublattices. An anisotropic NN bond couples these local responses, generating the unequal crossed displacement correlations responsible for circulation even when the antisymmetric interaction vanishes. Relaxation through anisotropic O--Se links converts the same environment contrast into an antisymmetric intersite coupling.


\section{
Phonon angular momentum}
\label{sec:AM}
The geometry of the lattice is encoded in the elastic couplings. In particular, the orthogonal local environments of $A$ and $B$ sites couple NN atomic displacements in a directionally dependent manner. 
To examine how these directional
interactions govern angular momentum exchange, we introduce the
local phonon angular momentum associated with atomic motion about
each equilibrium position~\cite{zhang2014angular,VonsovskiiSvirskii1962,Levine1962}:
\begin{equation}
\hat L_i=(\hat{\mathbf u}_i\times\hat{\mathbf p}_i)_z.
\label{Liang}
\end{equation}
The associated total phonon angular momentum
\begin{equation}
\hat L=\sum_i\hat L_i,
\end{equation}
generates a common rotation of all displacement and momentum vectors: $\hat{\mathbf{u}}_i\mapsto R_z(\theta)\hat{\mathbf{u}}_i$, and similarly for $\hat{\mathbf{p}}_i$ with $R_z(\theta)$ the rotation matrix about $z$ by $\theta$. 
Unlike a crystal rotation, which also acts on the lattice and may
exchange sites, this internal rotation acts only on the vibrational
degrees of freedom: its separation from rigid-body motion is
central to microscopic treatments of total angular momentum conservation~\cite{Ruckriegel2020}. The dependence of the elastic couplings on the lattice geometry generally breaks this continuous internal symmetry in crystals, allowing finite angular momentum fluctuations in the phonon vacuum even when the mean vanishes, as shown in Ref.~\cite{YiWilliamsFlebus2026}.

To separate angular momentum-conserving interactions from those that generate crystalline restoring torques, we decompose the elastic tensor into its isotropic and traceless parts, and let $\phi_{ij}$ denote the angle of the equilibrium bond direction relative to the $x$ axis, i.e., $\mathbf e_{ij}=(\cos\phi_{ij},\,\sin\phi_{ij})^{T}$.
The elastic tensor~\eqref{eq:elastic-tensor} takes the form
\begin{align}
K_{ij}
&=
\frac{\Gamma_{ij}
+\Gamma_{ij,T}
}{2}\mathbb I_2
+
\frac{\Gamma_{ij}
-\Gamma_{ij,T}
}{2}Q(\phi_{ij}),
\\
Q(\phi)
&=
\begin{pmatrix}
\cos 2\phi & \sin 2\phi\\
\sin 2\phi & -\cos 2\phi
\end{pmatrix},
\end{align}
where the traceless tensor $Q(\phi)$ has eigenvalue $+1$ along the bond and $-1$ perpendicular to it. 

We write the Hamiltonian as
$\hat H=\sum_{\langle ij\rangle}\hat h_{ij}$,
where each undirected NN or NNN bond is counted once.
The bond energy $\hat h_{ij}$ includes the elastic interaction
and an equal share of the kinetic energy at each endpoint.
The complete bond energy can be decomposed as
\begin{equation}
\hat h_{ij}=\hat h^{\mathrm{inv}}_{ij}+\hat h^\tau_{ij},
\end{equation}
with 
\begin{align}
\hat h^{\mathrm{inv}}_{ij}
&=
\frac{\Gamma_{ij}
+\Gamma_{ij,T}
}{4}
\left|\hat{\mathbf u}_j-\hat{\mathbf u}_i\right|^2
-d_{ij}(\hat{\mathbf u}_i\times\hat{\mathbf u}_j)_z \nonumber
\\
&+\frac{1}{2z}(\hat{\mathbf{p}}_i^2+\hat{\mathbf{p}}_j^2),
\label{Hinv}
\end{align}
and
\begin{align}
\hat h^\tau_{ij}
=&
\frac{\Gamma_{ij}
-\Gamma_{ij,T}
}{4}
(\hat{\mathbf u}_j-\hat{\mathbf u}_i)^T
Q(\phi_{ij})
(\hat{\mathbf u}_j-\hat{\mathbf u}_i),
\label{hTeq}
\end{align}
where
\begin{equation}
d_{ij}=
\begin{cases}
\eta_{ij}\lambda,
& i\in A,\ j\in B \quad\text{on NN bonds},\\
0,
& \langle ij\rangle\quad\text{on NNN bonds}
\end{cases}
\end{equation}
subject to $d_{ji} = -d_{ij}$
and with $z=8$ being the coordination number. The tensor $Q$ is unchanged under bond reversal, since $Q(\phi+\pi)=Q(\phi)$.
Both the squared displacement difference and the cross product are invariant under a common 
internal rotation.  That is,
\begin{equation}
[\hat h^{\mathrm{inv}}_{ij},\hat L_i+\hat L_j]=0.
\label{365eq}
\end{equation}
The onsite contributions to $\hat h_{ij}^\text{inv}$ in Eq.\,\eqref{Hinv} individually commute with $\hat{L}_i$ and $\hat{L}_j$, while Eq.\,\eqref{365eq} implies that the intersite terms can change $\hat L_i$ and $\hat L_j$ by equal and opposite amounts. This exchange of angular momentum between the two sites provides the basis for the bond-current operator defined in Sec.~\ref{sec:Current}. By contrast, the anisotropic term $\hat h^\tau_{ij}$ in
Eq.~\eqref{hTeq} depends on the angle between the relative
displacement and the fixed bond direction. This angular dependence
generally breaks internal rotational symmetry and gives rise
to crystalline torques. Mathematically, 
\begin{equation}
[\hat h_{ij}^\tau,\hat{L}_i+\hat{L}_j] \neq 0.
\end{equation}


A local circular representation makes the angular momentum
content of these interactions explicit. For $\sigma=\pm1$, we define 
\begin{equation}
\hat u_{i\sigma}
=
\frac{\hat u_{ix}-i\sigma\hat u_{iy}}{\sqrt{2}},
\qquad
\hat p_{i\sigma}
=
\frac{\hat p_{ix}-i\sigma\hat p_{iy}}{\sqrt{2}},
\label{eq:circular-components}
\end{equation}
where
$\hat u_{i\sigma}^{\dagger}=\hat u_{i,-\sigma}$ and
$\hat p_{i\sigma}^{\dagger}=\hat p_{i,-\sigma}$, and
\begin{equation}
[\hat u_{i\sigma},\hat p_{j\sigma'}^{\dagger}]
=
i\hbar\delta_{ij}\delta_{\sigma\sigma'},
\qquad
[\hat u_{i\sigma},\hat p_{j\sigma'}]
=
i\hbar\delta_{ij}\delta_{\sigma,-\sigma'}.
\end{equation}
In these variables, Eq.~\eqref{Liang} reads as
\begin{equation}
\hat L_i
=
i\left(
\hat u_{i,-}\hat p_{i,+}
-
\hat u_{i,+}\hat p_{i,-}
\right),
\label{eq:angular-momentum-circular}
\end{equation}
where its action on either circular component is
\begin{equation}
[\hat L_i,\hat u_{i\sigma}]
=
-\sigma\hbar\hat u_{i\sigma},
\qquad
[\hat L_i,\hat p_{i\sigma}]
=
-\sigma\hbar\hat p_{i\sigma}.
\label{eq:circular-angular-momentum-charge}
\end{equation}
 Viewing $\hat L_i+\hat L_j$ as a bond-local $U(1)$
(angular momentum) charge generating a common internal rotation
about $z$, we see from Eq.~\eqref{eq:circular-angular-momentum-charge} that $\hat u_{i\sigma}^{\dagger}$ carries charge
$\sigma\hbar$, while its adjoint $\hat u_{i\sigma}$ carries
$-\sigma\hbar$, and similarly for $\hat p_{i\sigma}^{\dagger}$
and $\hat p_{i\sigma}$.

The product $\hat u_{i\sigma}^{\dagger}\hat u_{j\sigma}$ carries
zero net bond-local $U(1)$ charge and therefore conserves
$\hat L_i+\hat L_j$. For a bond oriented from $i$ to $j$,
the identity
\begin{equation}
\hat u_{i\sigma}^{\dagger}\hat u_{j\sigma}
=
\frac{1}{2}
\left[
\hat{\mathbf u}_i\cdot\hat{\mathbf u}_j
-
i\sigma(\hat{\mathbf u}_i\times\hat{\mathbf u}_j)_z
\right]
\label{eq:circular-bond-identity}
\end{equation}
expresses the symmetric and antisymmetric bond interactions
in terms of this same charge-neutral bilinear.
By introducing
\begin{equation}
t_{ij,\sigma}
=
\frac{\Gamma_{ij}+\Gamma_{ij,T}}{2}
+i\sigma d_{ij},
\,\,\,\,
t_{ij,\sigma}^*=t_{ij,-\sigma} = t_{ji,\sigma},
\label{eq:tdef}
\end{equation}
 we can rewrite Eq.~\eqref{Hinv} as
\begin{equation}
\begin{aligned}
\hat h^{\mathrm{inv}}_{ij}
=-&\sum_{\sigma=\pm 1}\Bigg[
t_{ij,\sigma}
\hat u_{i\sigma}^{\dagger}\hat u_{j\sigma}
\\
-&\sum_{s=i,j}\left(
\frac{\Gamma_{ij}+\Gamma_{ij,T}}{4}\hat u_{s\sigma}^{\dagger}\hat u_{s\sigma} + \frac{\hat{p}_{s\sigma}^\dagger \hat{p}_{s\sigma}}{2z}\right)
\Bigg].
\end{aligned}
\label{inveq}
\end{equation}
Every term in Eq.~\eqref{inveq} carries zero bond-local $U(1)$
charge, consistent with Eq.~\eqref{365eq}.
The antisymmetric coupling enters the conserving exchange as an
imaginary contribution with opposite signs for the two circular
components. Consequently, the NN amplitudes in
Eq.~\eqref{eq:tdef} satisfy $t_{ij,+}=t_{ij,-}^{*}$ and carry
opposite phases, while the NNN amplitudes remain real because
these bonds contain only symmetric elastic couplings.

In the same representation, the anisotropic bond energy~\eqref{hTeq} reads as
\begin{equation}
\begin{aligned}
\hat h^\tau_{ij}
={}&
\frac{\Gamma_{ij}-\Gamma_{ij,T}}{4}
\sum_{\sigma=\pm1}e^{2i\sigma\phi_{ij}}
\left(\hat u_{j\sigma}-\hat u_{i\sigma}\right)^2.
\end{aligned}
\label{eq:anisotropic-bond-circular}
\end{equation}
Each squared term carries bond-local $U(1)$ charge
$-2\sigma\hbar$, as follows from
Eq.~\eqref{eq:circular-angular-momentum-charge}.
The stiffness contrast $\Gamma_{ij}-\Gamma_{ij,T}$ sets the
coupling strength, while the phase $e^{2i\sigma\phi_{ij}}$
encodes the bond orientation: it is real for axial NNN bonds
and imaginary for diagonal NN bonds.
We note that Eqs.~\eqref{inveq} and \eqref{eq:anisotropic-bond-circular}
are exact expressions of the original interactions in the
circular basis and involve no approximation.

Further insights arise from decomposing the total anisotropic energy into onsite and intersite  contributions:
\begin{equation}
\hat H^\tau
=
\sum_{\langle ij\rangle}\hat h^\tau_{ij}
=
\hat H^\tau_{\mathrm{on}}
+
\hat H^\tau_{\mathrm{inter}},
\label{eq:HT-on-inter-decomposition}
\end{equation}
where each undirected bond $\langle ij\rangle$ is counted once.
The onsite contribution $\hat H^\tau_{\mathrm{on}}$ describes
the directional restoring force on a displaced atom when its
neighbors are held fixed. The intersite contribution
$\hat H^\tau_{\mathrm{inter}}$ couples neighboring displacements,
describing how the motion of one atom modifies the force on another
and correlates their motion.
The orientation-dependent phase factors in Eq.\, \eqref{eq:anisotropic-bond-circular} force the NN contributions to $\hat H_\text{on}^\tau$ to vanish. 
Thus, $\hat H^\tau_\text{on}$ consists only of NNN contributions: 
\begin{equation}
\begin{aligned}
\hat H^\tau_{\mathrm{on}}
=&
\frac{
(\Gamma_2-\Gamma_{2T})
-
(\Gamma'_2-\Gamma'_{2T})
}{2}
\\
\times&\left(
\sum_{i\in A}-\sum_{i\in B}
\right)
\left(
\hat u_{i+}^{\,2}+\hat u_{i-}^{\,2}
\right).
\label{627}
\end{aligned}
\end{equation}
The opposite signs on $A$ and $B$ 
reflect the mutually orthogonal local environments. 
Expressing Eq.~\eqref{627} in Cartesian coordinates makes this explicit:
\begin{equation}
\begin{aligned}
\hat H^\tau_{\rm on}
=&
\frac{
(\Gamma_2-\Gamma_{2T})-(\Gamma^\prime_2-\Gamma^\prime_{2T})
}{2}
\\
\times & \left[
\sum_{i\in A}
\left(
\hat u_{ix}^{\,2}-\hat u_{iy}^{\,2}
\right) -
\sum_{i\in B}
\left(
\hat u_{ix}^{\,2}-\hat u_{iy}^{\,2}
\right)
\right].
\end{aligned}
\label{eq:HT-on}
\end{equation}
For $\Gamma_2-\Gamma_{2T}>\Gamma^\prime_2-\Gamma^\prime_{2T}$, the local restoring forces are stronger along $x$ at $A$ sites and along $y$ at $B$ sites, as illustrated in Fig.~\ref{toymodel}.
The NN bonds contribute to the intersite term:
\begin{equation}
\begin{aligned}
\hat H^\tau_{\mathrm{inter}}
&=
-\frac{\Gamma_1-\Gamma_{1T}}{2}
\sum_{\langle ij\rangle}
\sin 2\phi_{ij}
\left(
\hat u_{ix}\hat u_{jy}
+
\hat u_{iy}\hat u_{jx}
\right)+\cdots,
\end{aligned}
\label{Ht:inter}
\end{equation}
where the sum is over only NN bonds and the $\cdots$ represent the remaining NNN contribution.
Thus, Eqs.~\eqref{eq:HT-on}--\eqref{Ht:inter} show that the NNN stiffness difference $(\Gamma_2-\Gamma_{2T})-(\Gamma^\prime_2-\Gamma^\prime_{2T})$ establishes the local directional imbalance, while the NN bonds provide an anisotropic intersite coupling $\Gamma_1-\Gamma_{1T}$ that acts with equal strength in the locally hard--hard and soft--soft channels.

\vspace{1cm}

\section{Bond Currents,
Torques, \\ and  symmetry constraints}
\label{sec:Current}

We now examine how these interactions enter the local angular momentum balance.
The preceding decomposition separates the rotationally invariant part, which transfers angular momentum between sites through equal and opposite changes at the two endpoints, from the anisotropic part, which generates crystalline restoring torques.
To define the current associated with the conserving exchange of angular momentum, we introduce a fictitious $U(1)$ gauge field $A_{ij}$ on each bond $\langle ij\rangle$ within the invariant sector, following gauge and exchange-phase descriptions of equilibrium spin currents~\cite{Tokatly2008Equilibrium,SchuetzKollarKopietz2003}:
\begin{equation}
    t_{ij,\sigma} \mapsto e^{i\sigma A_{ij}} t_{ij,\sigma},
\end{equation}
where $A_{ij}=-A_{ji}$. Under a site-local $U(1)$ rotation $\hat{\mathbf{u}}_i\mapsto R_z(\theta_i) \hat{\mathbf{u}}_i$, the transformation 
$
A_{ij}\mapsto A_{ij} -\theta_i+\theta_j
$
leaves the bond energy unchanged.
This fictitious gauge field allows us to compute the bond-local current as
\begin{equation}
\begin{aligned}
\hat J_{i\to j}
&=\left.
\frac{\partial\hat h_{ij}^{\rm inv}(A_{ij})}
{\partial A_{ij}}
\right|_{A_{ij}=0}
\\
&= - i\left( t_{ij,+} \hat{u}_{i+}^\dagger \hat{u}_{j+}-t_{ij,-} \hat{u}_{i-}^\dagger \hat{u}_{j-}\right)
\\
&=-\frac{\Gamma_{ij}+\Gamma_{ij,T}}{2}
(\bu_i\times\bu_j)_z
+d_{ij}\,\bu_i\cdot\bu_j,
\end{aligned}
\label{source:current}
\end{equation}
with $\hat{J}_{j\to i} = -\hat{J}_{i\to j}$.
Equivalently, 
\begin{align}
   \hat J_{i\to j} &=-\frac{i}{\hbar}
[\hat h_{ij}^{\rm inv},\hat L_i]  = \frac{i}{\hbar}
[\hat h_{ij}^{\rm inv},\hat L_j],
\end{align} 
so the transfer changes the endpoint angular momenta by equal
and opposite amounts. Thus, we find
\begin{equation}
\dot{\hat L}_i+\sum_j\hat J_{i\to j}=\hat\tau_i,
\qquad
\hat\tau_i=\frac{i}{\hbar}[\hat H^{\tau},\hat L_i],
\label{eq:balance}
\end{equation}
where the sum is over all the sites $j$ connected to  the site $i$ by a bond. Equation~\eqref{eq:balance} shows that the invariant part of the Hamiltonian contributes locally-conserving currents while the anisotropic term yields nonconserving torques, i.e., sources and sinks of angular momentum. Both contributions must be considered together, as emphasized in the spin-current literature~\cite{Rashba2003,Shi2006SpinCurrent}.
Real-space phonon angular momentum currents have also been formulated for thermally driven Hall transport~\cite{BustamanteLopez2026Atomistic}. The present construction isolates exchange through the rotationally invariant bond interaction and retains crystalline restoring torques explicitly.

Crystal symmetry constrains the expectation values of the bond currents $\hat{J}_{i\to j}$ and local torques $\hat{\tau}_i$. We derive these constraints for any density operator $\hat\rho$ invariant under the crystal space group, including the phonon vacuum and thermal Gibbs states, with expectation values denoted by $\langle\hat O\rangle_\rho=\operatorname{Tr}(\hat\rho\hat O)$.
For an in-plane spatial symmetry $g$ represented by an in-plane orthogonal matrix $O_g$, the displacement, momentum, and axial angular momentum operators transform as
\begin{align}
\hat{U}_g \hat{\mathbf{u}}_i\hat U_g^{-1} &= O_g \hat{\mathbf{u}}_{g(i)},
\\
\hat{U}_g \hat{\mathbf{p}}_i\hat U_g^{-1} &= O_g \hat{\mathbf{p}}_{g(i)},
\\
\hat{U}_g\hat L_i\hat U_g^{-1} & = 
    \det(O_g)\hat L_{g(i)},
\end{align}
respectively, where $g(i)$ is defined implicitly by $\mathbf{R}_{g(i)} = O_g^{-1}\mathbf{R}_i$.
The bond currents therefore satisfy
\begin{equation}
\hat U_g\hat J_{i\to j}\hat U_g^{-1}
=
\det(O_g)\hat J_{g(i)\to g(j)},
\qquad
\hat J_{j\to i}=-\hat J_{i\to j}.
\label{eq:current-spatial-symmetry}
\end{equation}
In a symmetry preserving state, this relation constrains the mean bond currents.

The $C_2$ symmetry about an NNN bond midpoint reverses $\hat J_{i\to j}$ [Eq.~\eqref{eq:current-spatial-symmetry}], forcing its expectation value to vanish in any invariant state $\hat{\rho}$:
\begin{equation}
\langle\hat J_{i\to j}\rangle_\rho=0,
\qquad
ij\text{ an NNN bond},
\label{eq:nnn-current-zero}
\end{equation}
so neither $AA$ nor $BB$ bonds carry a mean equilibrium current.
A diagonal nearest-neighbor (NN) bond lacks midpoint $C_2$ symmetry, but reflection across its perpendicular bisector remains a crystal symmetry. This reflection exchanges the endpoints and reverses the axial angular momentum, leaving the bond-current operator unchanged. A nonzero NN current is thus fully compatible with crystal symmetry.
Nevertheless, symmetry constrains its spatial pattern: axial mirror operations map the four NN bonds incident on any given site onto one another, fixing the relative signs of their currents. Specifically, for $A$ and $B$ sites connected by a bond vector
$a(\eta_x,\eta_y)/2$ directed from $A$ to $B$, with $\eta_x,\eta_y\in\{\pm1\}$, we find
\begin{equation}
\langle\hat J_{A\to B,\eta_x\eta_y}\rangle_\rho
=
\eta_x\eta_y I_\rho,
\qquad
I_\rho\equiv
\langle\hat J_{A\to B,++}\rangle_\rho.
\label{eq:nn-current-pattern}
\end{equation}
Consequently, the total current flowing from any site to its four nearest neighbors vanishes.
Time reversal imposes no additional constraint on $I_\rho$: angular momentum current is time-reversal even, whereas angular momentum itself is odd.
The crystalline torques identified by Eq.~\eqref{eq:balance} transform as
\begin{equation}
\hat U_g\hat\tau_i\hat U_g^{-1}
=
\det(O_g)\hat\tau_{g(i)},
\label{eq:torque-spatial-symmetry}
\end{equation}
and therefore satisfy analogous symmetry constraints.
An axial mirror through site $i$ leaves the site fixed but reverses its torque, forcing
\begin{equation}\label{eq: tauvanish}
\langle\hat\tau_i\rangle_\rho=0.
\end{equation}
Taking the expectation value of Eq.~\eqref{eq:balance} yields an equivalent result, as the onsite and intersite parts of $\hat{H}^{\tau}$ are separately invariant under this mirror operation.
\begin{figure*}[t]
\includegraphics[width=\textwidth]{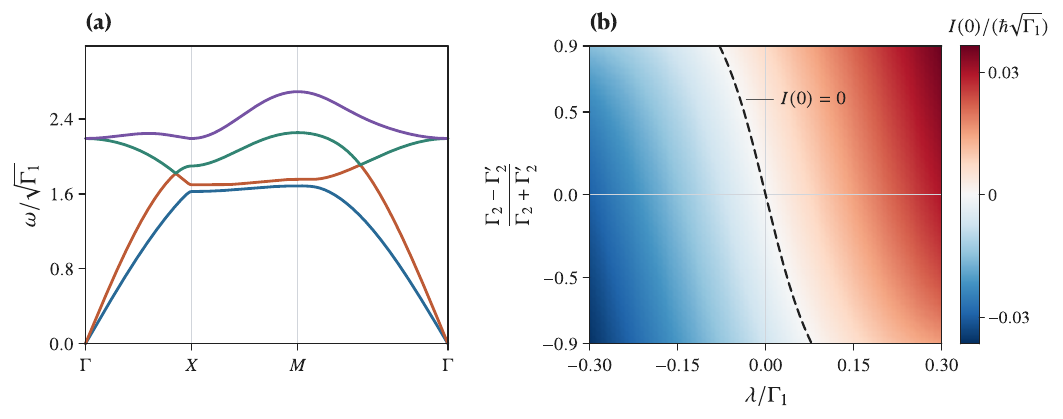}
\caption{Phonon spectrum and equilibrium angular momentum circulation.
(a) Phonon dispersion obtained from Eq.~\eqref{eq:H} along the
high-symmetry path $\Gamma-X-M-\Gamma$ where
$\Gamma=(0,0)$, $X=(\pi/a,0)$, and $M=(\pi/a,\pi/a)$, with frequencies
expressed in units of $\sqrt{\Gamma_1}$, for
$\lambda/\Gamma_1=0.15$ and
$(\Gamma_2-\Gamma_2')/(\Gamma_2+\Gamma_2')=1/3$.
(b) Ground-state current on the reference nearest-neighbor bond,
$I(0)=\langle\hat{J}_{A\to B,++}\rangle_0$, evaluated from
Eq.~\eqref{curre} at $T=0$ and expressed in units of
$\hbar\sqrt{\Gamma_1}$, as a function of the antisymmetric
coupling $\lambda/\Gamma_1$ and the elastic-environment contrast
$(\Gamma_2-\Gamma_2')/(\Gamma_2+\Gamma_2')$.
The dashed line indicates where $I(0)=0$, across which
the circulation reverses. A finite current persists at
$\lambda=0$ when $\Gamma_2\neq\Gamma_2'$.
Both panels use $\Gamma_{1T}/\Gamma_1=0.2$,
$(\Gamma_2+\Gamma_2')/\Gamma_1=0.9$, and
$\Gamma_{2T}/\Gamma_2=\Gamma_{2T}'/\Gamma_2'=0.2$.}
\label{fig:vacuum}
\end{figure*}
Hence, by defining the torque contributions from either family of interactions
$\hat\tau_i^\alpha=(i/\hbar)
[\hat H^{\tau}_\alpha,\hat L_i]$ for
$\alpha\in\{{\rm on},{\rm inter}\}$, we find
\begin{equation}
\langle\hat\tau_i^{\rm on}\rangle_\rho
=
\langle\hat\tau_i^{\rm inter}\rangle_\rho
=0.
\label{eq:site-torque-zero}
\end{equation}
We emphasize that these are constraints on expectation values, not operator identities: the anisotropic interactions remain present in the dynamics and, as we will show next, shape the equilibrium correlations.

To connect the site-local torque balance in Eq.~\eqref{eq: tauvanish} to the restoring response of a single NN bond, we define the endpoint torques generated by the bond energy in Eq.~\eqref{eq:anisotropic-bond-circular}.
For $i\in A$ and $j\in B$, with the bond vector specified in Eq.~\eqref{eq: bondvec}, these torques are
\begin{equation}
\hat\tau_{A,\eta}
=
\frac{i}{\hbar}[\hat h^\tau_{AB},\hat L_A],
\qquad
\hat\tau_{B,\eta}
=
\frac{i}{\hbar}[\hat h^\tau_{AB},\hat L_B],
\label{eq:bond-endpoint-torques}
\end{equation}
where $\eta=\eta_x\eta_y$ denotes the bond sign factor introduced below Eq.~\eqref{eq: bondvec}.
A direct calculation yields
\begin{equation}
\begin{aligned}
\hat{\tau}_{A,\eta}
&=
\frac{\eta\left(\Gamma_1-\Gamma_{1T}\right)}{2}
\left[
\hat{u}_{Ax}\hat{u}_{Bx}
-\hat{u}_{Ay}\hat{u}_{By}
-\hat{u}_{Ax}^{2}
+\hat{u}_{Ay}^{2}
\right],
\\
\hat{\tau}_{B,\eta}
&=
\frac{\eta\left(\Gamma_1-\Gamma_{1T}\right)}{2}
\left[
\hat{u}_{Ax}\hat{u}_{Bx}
-\hat{u}_{Ay}\hat{u}_{By}
-\hat{u}_{Bx}^{2}
+\hat{u}_{By}^{2}
\right].
\end{aligned}
\end{equation}
For the selected NN bond, reflection across its perpendicular bisector maps $\hat\tau_{A,\eta}$ to $-\hat\tau_{B,\eta}$.
Consequently, in a symmetry-preserving state,
\begin{equation}
\langle\hat\tau_{A,\eta}\rangle_\rho
=
-\langle\hat\tau_{B,\eta}\rangle_\rho,
\label{eq:endpoint-torque-symmetry}
\end{equation}
without requiring either mean to vanish separately.
These bond-resolved responses cancel upon summing over the symmetry-related NN bonds incident on a given site.
For an axial NNN bond, reflection in the bond axis instead leaves both endpoints fixed and reverses each endpoint torque, forcing each mean torque to vanish.

The contribution of the selected $AB$ bond to the continuity equation~\eqref{eq:balance} is
\begin{equation}
\begin{aligned}
\left.\dot{\hat L}_A\right|_{AB}
+\hat J_{A\to B}
&=\hat\tau_{A,\eta},
\\
\left.\dot{\hat L}_B\right|_{AB}
-\hat J_{A\to B}
&=\hat\tau_{B,\eta},
\end{aligned}
\label{eq:bond-balance}
\end{equation}
where
$\left.\dot{\hat L}_s\right|_{AB}
=(i/\hbar)[\hat h^{\rm inv}_{AB}+\hat h^\tau_{AB},\hat L_s]$,
with $s\in\{A,B\}$.
The invariant interaction changes the endpoint angular momenta at equal and opposite rates, whereas the anisotropic endpoint torques need not vanish.
We define the bond-resolved restoring response as
\begin{align}
\hat T_\eta
&=\frac{\hat\tau_{B,\eta}-\hat\tau_{A,\eta}}{2}
\notag\\
&=\frac{\eta\left(\Gamma_1-\Gamma_{1T}\right)}{4}
\left(\hat u_{Ax}^2-\hat u_{Ay}^2-\hat u_{Bx}^2+\hat u_{By}^2\right).
\label{eq:antisymmetric-endpoint-torque}
\end{align}
The equal endpoint torques generated by the anisotropic intersite coupling cancel in this combination, leaving the difference between the single-bond onsite torques.
Equation~\eqref{eq:bond-balance} then gives
\begin{equation}
\left.
\frac{d}{dt}\left(\frac{\hat L_B-\hat L_A}{2}\right)
\right|_{AB}
-
\hat J_{A\to B} = \hat T_\eta.
\label{eq:relative-bond-balance}
\end{equation}
Equation~\eqref{eq:relative-bond-balance} describes how the $AB$ bond interaction changes the staggered angular momentum $\hat L_{\mathrm{st}}=\hat L_A-\hat L_B$: the current transfers angular momentum between the two sites, while $\hat T_\eta$ accounts for the difference between their crystalline restoring torques.


\section{ Equilibrium Circulation}
\label{sec:EC}

We proceed by evaluating the symmetry-allowed current amplitude and the bond-resolved restoring response in the phonon vacuum; their finite-temperature generalization follows directly from the same formalism.
Because the bond-current operator is bilinear in the displacements, its equilibrium expectation value is determined entirely by the equal-time displacement covariance.
For the harmonic Hamiltonian, this covariance is obtained exactly from the normal modes.

Using the Fourier convention in Eq.~\eqref{eq:fourier-convention}, we construct the mass-weighted dynamical matrix $D(\mathbf k)$ from the Hamiltonian~\eqref{eq:H}.
Its explicit form is given in Appendix~\ref{app:DynamicalMatrix}.
The phonon angular frequencies and normalized polarization vectors follow from
\begin{equation}
D(\mathbf k)\boldsymbol\varepsilon_{\mathbf k\nu}
=
\omega_{\mathbf k\nu}^{2}\boldsymbol\varepsilon_{\mathbf k\nu},
\qquad
\boldsymbol\varepsilon_{\mathbf k\nu}^{\dagger}
\boldsymbol\varepsilon_{\mathbf k\nu'}
=
\delta_{\nu\nu'},
\label{PhononEigenproblem}
\end{equation}
where
$\boldsymbol\varepsilon_{\mathbf k\nu}
=(\varepsilon_{Ax,\mathbf k\nu},
\varepsilon_{Ay,\mathbf k\nu},
\varepsilon_{Bx,\mathbf k\nu},
\varepsilon_{By,\mathbf k\nu})^{T}$
specifies the displacement polarization on the two sublattices.
Here $\mathbf k$ is the crystal wave vector, $\nu=1,\ldots,4$ labels the phonon branches, and $\omega_{\mathbf k\nu}$ is the corresponding angular frequency shown in Fig.\,\ref{fig:vacuum}(a).
We consider a finite  crystal of $N_c$ unit cells under periodic boundary conditions in a mechanically stable regime, with no zero modes other than the two rigid translations.
To restrict the calculation to internal vibrations, we omit these translations from the primed sums, so that all retained modes satisfy $\omega_{\mathbf k\nu}>0$.
The wave-vector sum runs over the $N_c$ allowed points in the first Brillouin zone, with the positive-frequency optical modes at $\mathbf k=0$ retained.

The equilibrium displacement covariance between sites $i$ and $j$ can be written as~\cite{Glauber1955}
\begin{equation}
\begin{aligned}
\langle\hat u_{i\alpha}\hat u_{j\beta}\rangle_T
={}&
\frac{\hbar}{2N_c}
\sum_{\mathbf k\nu}^{\prime}
g_T(\omega_{\mathbf k\nu})
\operatorname{Re}\!\left[
e^{i\mathbf k\cdot(\mathbf R_j-\mathbf R_i)}
\varepsilon_{s\alpha,\mathbf k\nu}^{*}
\varepsilon_{s'\beta,\mathbf k\nu}
\right],
\end{aligned}
\label{eq:displacement-covariance}
\end{equation}
where $\alpha,\beta=x,y$, and $\langle\cdots\rangle_T$
represents the equilibrium expectation at temperature $T$, with
$T=0$ denoting the phonon vacuum.
The thermal weight is defined as 
\begin{equation}
g_T(\omega)
=
\frac{1}{\omega}
\coth\!\left(\frac{\hbar\omega}{2k_BT}\right),
\qquad
g_0(\omega)=\frac{1}{\omega},
\label{eq:thermal-weight}
\end{equation}
where $k_B$ is the
Boltzmann constant, and
the weight $g_T$ includes both zero-point and thermal
fluctuations.
Taking the thermal expectation value of Eq.~\eqref{source:current}, we express the current $I(T)\equiv I_\rho$ as the sum of $I^{\rm cross}(T)$, which probes the imbalance between crossed displacement correlations, and $I^\lambda(T)$, which probes the relative alignment of neighboring displacements through the antisymmetric bond interaction. The crossed-displacement contribution is
\begin{equation}
I^{\mathrm{cross}}(T)
=
-\frac{\Gamma_1+\Gamma_{1T}}{2}
\left\langle(\hat{\mathbf u}_A\times\hat{\mathbf u}_B)_z\right\rangle_T.
\label{eq:cross-current-covariance}
\end{equation}
A nonzero $I^{\mathrm{cross}}(T)$ therefore requires unequal crossed displacement correlations:
\begin{align}
\langle\hat u_{Ax}\hat u_{By}\rangle_T
\neq
\langle\hat u_{Ay}\hat u_{Bx}\rangle_T.
\label{eq: unequalcross}
\end{align}
Appendix~\ref{app:DynamicalMatrix} evaluates this difference
at $\lambda=0$ for a mechanically stable lattice with
$\Gamma_1,\Gamma_{1T},\Gamma_2,\Gamma'_2>0$ and
$\Gamma_{2T}=\rho\Gamma_2$,
$\Gamma'_{2T}=\rho\Gamma'_2$, where $0\leq \rho<1$.
Under these assumptions, Eq.~\eqref{eq:sign} shows that
$I^{\mathrm{cross}}(T)$ is nonzero for all $T\geq0$
whenever both $\Gamma_1\neq\Gamma_{1T}$ and
$\Gamma_2\neq\Gamma'_2$.
At finite $\lambda$, the antisymmetric bond interaction contributes through the scalar-product correlator 
\begin{equation}
I^\lambda(T)
\equiv
\lambda\langle\hat{\mathbf u}_A\cdot\hat{\mathbf u}_B\rangle_T.
\label{eq:lambda-current-covariance}
\end{equation}
Combining Eqs.~\eqref{eq:cross-current-covariance} and
\eqref{eq:lambda-current-covariance} gives the full mean current:
\begin{widetext}
\begin{equation}
\begin{aligned}
I(T)
&\equiv\langle\hat J_{A\to B,++}\rangle_T
=\frac{\hbar}{2N_c}
\sum_{\mathbf k\nu}^{\prime}
g_T(\omega_{\mathbf k\nu})
\operatorname{Re}\Biggl\{
e^{i\mathbf k\cdot\boldsymbol\delta_{++}}\Biggl[
\lambda\,
\boldsymbol\varepsilon_{A,\mathbf k\nu}^{*}
\cdot\boldsymbol\varepsilon_{B,\mathbf k\nu}
-\frac{\Gamma_1+\Gamma_{1T}}{2}
\left(
\boldsymbol\varepsilon_{A,\mathbf k\nu}^{*}
\times\boldsymbol\varepsilon_{B,\mathbf k\nu}
\right)_z
\Biggr]\Biggr\},
\end{aligned}
\label{curre}
\end{equation}
\end{widetext}
where $
\boldsymbol\delta_{++}\equiv\mathbf R_B-\mathbf R_A
=\frac{a}{2}(1,1)
$, and $\boldsymbol\varepsilon_{s,\mathbf k\nu}
\equiv(\varepsilon_{sx,\mathbf k\nu},
  \varepsilon_{sy,\mathbf k\nu})^T$
with $s=A,B$.
Equations~\eqref{eq:nnn-current-zero} and
\eqref{eq:nn-current-pattern} fix the full spatial pattern once
$I(T)$ is known: all NNN mean currents vanish, while the NN
currents have equal magnitude and signs set by $\eta_x\eta_y$.
For $I(T)>0$ ($I(T)<0$), the resulting arrows circulate clockwise (counterclockwise) around
O-centered diamond plaquettes and counterclockwise (clockwise) around
Se-centered plaquettes, as shown in Fig.~\ref{fig:model}(a). Each diamond is bounded by
four alternating $A$--$B$ bonds. Adjacent plaquettes share
a bond whose single current direction is consistent with these
opposite circulation senses.
Figure~\ref{fig:vacuum}(b) shows the current calculated from the complete
phonon spectrum using a common transverse ratio,
$\Gamma_{2T}=\rho\Gamma_2$ and $\Gamma^\prime_{2T}=\rho\Gamma^\prime_2$,
with $0\leq \rho<1$.
The heatmap exhibits a finite current both at $\lambda=0$, when the
two NNN environments differ, and at
$\Gamma_2=\Gamma^\prime_2$, when $\lambda\neq0$.
A zero-current contour separates regions of opposite current sign.

To identify the leading parameter dependence of these two responses,
we expand Eq.~\eqref{curre} at $T=0$ about
$\lambda=0$, $\Gamma_1=\Gamma_{1T}$, and $\Gamma_2=\Gamma^\prime_2$,
while holding $\Gamma_1+\Gamma_{1T}>0$,
$\Gamma_2+\Gamma^\prime_2>0$,  and the ratio $\rho$ fixed.
We assume that the reference spectrum is stable, with nonzero acoustic
sound velocities and no zero modes beyond the two rigid translations.
The deviations from this reference are small compared with the
NN stiffness scale, i.e.,
\[
\epsilon =
\frac{
\max\!\left(
|\lambda|,
|\Gamma_1-\Gamma_{1T}|,
|\Gamma_2-\Gamma^\prime_2|
\right)
}{
\Gamma_1+\Gamma_{1T}
}
\ll 1.
\]
Expanding to second order in these small deviations yields
\begin{equation}
\begin{aligned}
I^{\mathrm{cross}}(0)
={}&
-c_{\mathrm{cross}}
\frac{\hbar\lambda}{\sqrt{\Gamma_1+\Gamma_{1T}}}
\\
&+
c_{\mathrm{env}}
\frac{
\hbar(\Gamma_1-\Gamma_{1T})(\Gamma_2-\Gamma^\prime_2)
}{
(\Gamma_1+\Gamma_{1T})^{3/2}
}
\\
&+
O\!\left(
\hbar\sqrt{\Gamma_1+\Gamma_{1T}}\,\epsilon^3
\right),
\\[0.5em]
I^\lambda(0)
={}&
c_\lambda
\frac{\hbar\lambda}{\sqrt{\Gamma_1+\Gamma_{1T}}}
+
O\!\left(
\hbar\sqrt{\Gamma_1+\Gamma_{1T}}\,\epsilon^3
\right),
\end{aligned}
\label{eq:expansion}
\end{equation}
where the dimensionless coefficients $c_{\mathrm{cross}}$, $c_\lambda$,
and $c_{\mathrm{env}}$ are obtained by expanding the complete
mode sum in Eq.~\eqref{curre}.
At $\lambda=0$, the leading current in Eq.~\eqref{eq:expansion}
is proportional to
$(\Gamma_1-\Gamma_{1T})(\Gamma_2-\Gamma^\prime_2)$,
reflecting the interplay between anisotropic NN coupling and
local anisotropy illustrated in Fig.~\ref{toymodel}.
The local anisotropy arises from the NNN stiffness contrast
$\Gamma_2-\Gamma^\prime_2$ through the onsite potential defined in
Eq.~\eqref{eq:HT-on}: for $\Gamma_2>\Gamma^\prime_2$,
the locally hard directions are $A_x$ and $B_y$,
while $A_y$ and $B_x$ are soft.
The anisotropic NN bond in Eq.~\eqref{Ht:inter} couples these hard--hard and soft--soft
pairs with the same coefficient proportional to
$\Gamma_1-\Gamma_{1T}$, which allows their local anisotropy to be transferred as angular momentum current.  If $\Gamma_1=\Gamma_{1T}$, the Hamiltonian at $\lambda=0$
separates into independent $x$ and $y$ sectors, so both crossed
correlations vanish. If $\Gamma_2=\Gamma^\prime_2$, the two crossed channels
become equivalent and their contributions cancel, even when the
NN spring remains anisotropic.



A finite antisymmetric coupling $\lambda$ provides an additional route
to circulation. It favors a signed relative orientation of neighboring
displacement vectors while remaining invariant under their common
rotation. It enters the current operator explicitly through the
scalar-product term $I^\lambda$, and also modifies the ground-state crossed
correlations entering $I^{\mathrm{cross}}$.

Despite their different appearances, these two routes to equilibrium circulation share a common microscopic
origin in the parent elastic network: inequivalent local elastic
environments and spring forces with unequal longitudinal and transverse
stiffnesses. As shown in Eqs.~\eqref{eq:lambda}--\eqref{eq:Gamma},
making the two parent environments elastically equivalent
eliminates $\lambda$ and $\Gamma_2-\Gamma^\prime_2$. Mathematically,
\begin{equation}
\begin{gathered}
\begin{gathered}
k_{\mathrm O,L}=k_{\mathrm{Se},L}\\
k_{\mathrm O,T}=k_{\mathrm{Se},T}
\end{gathered}
\quad\Longrightarrow\quad
\begin{aligned}
\lambda &= 0\\
\Gamma_2-\Gamma_2^\prime &= 0
\end{aligned}
\end{gathered}
\end{equation}

A more general constraint relates equilibrium circulation to local
angular momentum fluctuations, independently of this particular parent
network. For a bond $ij$ with the current defined by
Eq.~\eqref{source:current}, the Robertson uncertainty
relation~\cite{Robertson1929}, applied to the invariant bond energy
in Eq.~\eqref{Hinv} and either endpoint angular momentum, gives
\begin{equation}
\begin{aligned}
\bigl|\langle\hat J_{i\to j}\rangle_0\bigr|
\le{}&
\frac{2}{\hbar}
\Biggl\{
\Bigl[
\langle(\hat h_{ij}^{\rm inv})^2\rangle_0
-\langle\hat h_{ij}^{\rm inv}\rangle_0^2
\Bigr]
\times
\langle\hat L_s^2\rangle_0
\Biggr\}^{1/2},
\end{aligned}
\label{eq:bond-current-upper-bound}
\end{equation}
with $s\in\{i,j\}$, and where 
 $\langle\hat L_s\rangle_0=0$ by time reversal. If the full
Hamiltonian conserves the angular momentum of either endpoint,
a nondegenerate vacuum has $\langle\hat L_s^2\rangle_0=0$;
Eq.~\eqref{eq:bond-current-upper-bound} then forces the mean bond
current to vanish. Internal isotropy of each spring alone does not
imply this local conservation condition: isotropic bonds can still
transfer angular momentum between their endpoints. Appendix~\ref{app:bond-current-bound}
explicitly derives this bound.

Importantly, neither mechanism requires spontaneous symmetry breaking or a
nonzero mean local angular momentum.
The inequivalent stiffness environments and the perpendicular local
restoring axes are already present in the crystal structure. Unequal crossed correlations and directed exchange are therefore
compatible with the full crystal symmetry, which organizes the bond
currents into opposite circulations on neighboring plaquettes.

The displacement covariance also determines the mean restoring
torques exerted by an individual bond. In a symmetry-preserving
equilibrium state, the fourfold rotation interchanges the
$x$ and $y$ displacement components on the two sublattices,
implying
\begin{equation}
\left\langle
\hat u_{Bx}^{\,2}-\hat u_{By}^{\,2}
\right\rangle_T
=
-\left\langle
\hat u_{Ax}^{\,2}-\hat u_{Ay}^{\,2}
\right\rangle_T.
\end{equation}
Using this relation, the bond-resolved restoring response defined in Eq.\,\eqref{eq:antisymmetric-endpoint-torque} becomes
\begin{equation}
\begin{aligned}
\left\langle\hat T_\eta\right\rangle_T
&=
\frac{\eta(\Gamma_1-\Gamma_{1T})}{2}
\left\langle
\hat u_{Ax}^{\,2}-\hat u_{Ay}^{\,2}
\right\rangle_T
\\
&=
\frac{\hbar\eta(\Gamma_1-\Gamma_{1T})}{4N_c}
\sum_{\mathbf k\nu}^{\prime}
g_T(\omega_{\mathbf k\nu})\\
&\quad\times
\left(
|\varepsilon_{Ax,\mathbf k\nu}|^2
-
|\varepsilon_{Ay,\mathbf k\nu}|^2
\right).
\end{aligned}
\label{eq:mean-endpoint-torque}
\end{equation}
The thermal weight $g_T$ is defined in Eq.~\eqref{eq:thermal-weight}, with
$g_0(\omega)=1/\omega$ recovering the phonon-vacuum result. Evaluating Eq.~\eqref{eq:mean-endpoint-torque} in the same limit used to derive Eq.~\eqref{eq:expansion}, we obtain
\begin{equation}
\begin{aligned}
\left\langle\hat T_\eta\right\rangle_0
={}&
-\frac{\eta q_{\rm env}}{2}\,
\frac{\hbar(\Gamma_1-\Gamma_{1T})(\Gamma_2-\Gamma^\prime_2)}
     {(\Gamma_1+\Gamma_{1T})^{3/2}}
\\[2pt]
&+
\frac{\eta q_{\rm mix}}{2}\,
\frac{\hbar\lambda(\Gamma_1-\Gamma_{1T})^2}
     {(\Gamma_1+\Gamma_{1T})^{5/2}}
\\[2pt]
&+
O\!\left(\hbar\sqrt{\Gamma_1+\Gamma_{1T}}\,\epsilon^4\right),
\end{aligned}
\label{eq:vacuum-torque-expansion}
\end{equation}
where  $q_{\rm env}$ and
$q_{\rm mix}$ are dimensionless response coefficients determined
by the reference phonon spectrum.

The exchange current and the torque half-difference probe distinct
components of the same equilibrium displacement covariance:
the former depends on intersite correlations, while the latter
measures the anisotropy of local fluctuations. Their dependence
on the same microscopic parameters therefore reflects a common
origin, without requiring their magnitudes to coincide.
At $\lambda=0$, the NNN contrast
$\Gamma_2-\Gamma^\prime_2$ produces
opposite local fluctuation anisotropies on the two sublattices.
The NN anisotropy $\Gamma_1-\Gamma_{1T}$ couples the orthogonal
displacement components across a bond and determines the
restoring torque exerted on these unequal fluctuations.
Both leading responses consequently involve the product
$(\Gamma_1-\Gamma_{1T})(\Gamma_2-\Gamma^\prime_2)$.
When the NNN environments become equivalent, this contribution
vanishes even if the individual NN springs remain anisotropic.
A finite $\lambda$ provides a distinct mechanism: it generates
an exchange current already at linear order, whereas its
leading contribution to the torque half-difference requires
two factors of NN anisotropy, one in the induced local
fluctuation anisotropy and one in the torque operator.

A finite mean restoring torque on an individual bond is
therefore compatible with vanishing total mean torque
at every site: the bond contributions cancel upon summing
over the local environment, while the equilibrium currents
satisfy $\sum_j\langle\hat J_{i\to j}\rangle_T=0$.

\section{Wilson Loops and Anisotropic Correlations}
\label{sec:WL}
Equilibrium circulation can often be understood through phases accumulated
around closed paths
~\cite{SchuetzKollarKopietz2003,FlynnFlebus2026}.
In the present lattice, however, angular momentum-conserving exchange
coexists with anisotropic interactions that fix preferred displacement
directions. These interactions shape the equilibrium correlations and
can thereby sustain a current even when their mean restoring torque
vanishes. Understanding the circulation therefore requires
distinguishing the geometry encoded in the conserving interactions from
the additional orientational structure of the full Hamiltonian.
To characterize the phase structure of the conserving exchange, we construct its gauge-invariant Wilson loops~\cite{Wilson1974}. 
For a given bond, the amplitude $t_{ij,\sigma}$ defined in Eq.~\eqref{eq:tdef} transforms as
\begin{align}
    t_{ij,\sigma}\mapsto e^{-i\sigma(\theta_i-\theta_j)}t_{ij,\sigma},
\end{align}
under the site-local rotation considered in Sec.\,\ref{sec:Current}. Viewing this as a local gauge transformation, the Wilson loops
\begin{align}
    W_\sigma({\mathcal{C}}) = \prod_{\langle ij\rangle\in\mathcal{C}} \frac{t_{ij,\sigma}}{|t_{ij,\sigma}|} = e^{i\Phi_\sigma(\mathcal{C})}
\end{align}
are gauge invariant for any directed closed path $\mathcal C$ on the lattice.
A nontrivial loop phase, $\Phi_\sigma(\mathcal C)\neq 2\pi n$, therefore rules out a choice of local frames in which all exchange phases vanish.

Consider the $\mathrm{La_2O_3Mn_2Se_2}$ realization of our effective lattice, i.e., Fig.\,\ref{fig:model}(b). For the counterclockwise paths $\mathcal C_{\rm Se}$ and $\mathcal C_{\rm O}$ around the Se- and O-centered plaquettes in Fig.~\ref{fig:model}(a), respectively, we obtain
\begin{equation}
\Phi_{\sigma}(\mathcal{C}_{\rm Se})
=
-\Phi_{\sigma}(\mathcal{C}_{\rm O})
=
4\sigma\arctan\!\left(\frac{2\lambda}{\Gamma_1+\Gamma_{1T}}\right),
\label{wilson:result}
\end{equation}
which are generically not integer multiples of $2\pi$. We observe that, within a fixed polarization sector $\sigma$, neighboring plaquettes have opposite phases. Similarly, the same plaquette hosts opposing phases between sectors. 
These nontrivial Wilson loop phases quantify frustration in the
angular momentum conserving sector: the relative orientations of displacement
fluctuations favored by different bonds cannot all be reconciled
around an elementary plaquette. This is directly analogous to
the hidden frustration of transverse spin fluctuations in
Ref.~\cite{FlynnFlebus2026}.

The limit $\lambda \to 0$ reveals the scope of this connection.
In this case, the Wilson phase on every plaquette is trivial.
Nevertheless, we have proven that the equilibrium circulation
can remain finite. This is in sharp contrast to the finding in
Ref.\,\cite{FlynnFlebus2026}, where both symmetric and
antisymmetric -- but still $U(1)$-conserving -- spin exchange
interactions (and thus a nontrivial Wilson flux) were necessary
for equilibrium spin currents.
The Wilson phases therefore capture only one mechanism for phonon circulation: the interplay between rotationally invariant symmetric and antisymmetric couplings.

The additional current is generated via the anisotropic interactions in $\hat H^\tau$, Eq.\,\eqref{eq:HT-on-inter-decomposition}. These interactions establish preferred displacement directions and, in the inequivalent local environments of the lattice, generate unequal crossed displacement correlations, see Eq.\,\eqref{eq: unequalcross}. 
Their imbalance gives a finite expectation value of the conserving bond current $I^\text{cross}(T)$ in Eq. ~\eqref{eq:cross-current-covariance}, evaluated in the equilibrium state of the full Hamiltonian. This contribution survives even in the absence of nontrivial exchange phases.

It is also essential to notice that, while crystal symmetries permit a scenario where the antisymmetric coupling $\lambda$ is nonzero but $\hat H^\tau=0$, the microscopic origin of $\lambda$ in the parent elastic network requires directional spring forces acting through inequivalent local environments [see Appendix~\ref{app:parent}]. Equivalently, in terms of symmetries, while the antisymmetric term itself is $U(1)$ symmetric, it originates microscopically from $U(1)$ symmetry breaking terms. If the longitudinal and transverse stiffnesses were to coincide on every parent bond (thus restoring $U(1)$ symmetry), $\lambda$ would vanish according to Eq.~\eqref{eq:effective_antisymmetric_coupling}.  

Directional elasticity thus provides a common microscopic origin
for both circulation mechanisms in the parent elastic network.
It also connects them to dynamical angular momentum fluctuations
in the phonon vacuum, which likewise rely on the noncommutativity
of the Hamiltonian with total phonon angular
momentum~\cite{YiWilliamsFlebus2026}.
It is worth mentioning that this constraint does not eliminate local endpoint
fluctuations or angular momentum exchange between sites,
as discussed in Appendix~\ref{app:bond-current-bound}. Circulation additionally requires spatially organized intersite
displacement correlations, revealing structure in the equilibrium
state that remains invisible to the local mean angular momentum.

\section{Discussion and Outlook}
\label{sec:discussion}
We have shown that directional elasticity in locally anisotropic crystal environments organizes quantum and thermal fluctuations into equilibrium angular-momentum circulation. The same microscopic ingredients can generate staggered Dzyaloshinskii–Moriya-like couplings between atomic displacements when internal ionic coordinates are eliminated. Both effects preserve time-reversal and inversion symmetry and arise without external driving, spontaneous symmetry breaking, or coupling to electronic or magnetic order.

In the effective lattice, symmetric, direction-dependent elastic interactions break the internal \(U(1)\) symmetry associated with a common rotation of the displacements relative to the crystal. Together with differently oriented local restoring axes, these interactions cause equally coupled displacement channels to develop correlations of unequal magnitude, which, in turn, gives rise to an equilibrium angular-momentum circulation. The same local geometry allows the relaxation of internal ionic coordinates to generate an antisymmetric intersite coupling from a parent network of symmetric spring tensors. This DMI-like interaction acts on the relative orientation of neighboring displacements, shaping collective vibrational modes and their spatial correlations. We demonstrate that this interaction provides an additional contribution to equilibrium angular-momentum circulation, associated with nontrivial Wilson loop phases in the angular-momentum-conserving exchange sector. Beyond its implications for equilibrium circulation, the emergence of this coupling calls for revisiting how first-principles force constants are reduced to effective lattice Hamiltonians, where restricting intersite interactions to symmetric forms can discard antisymmetric components. Resolving these components and their spatial pattern in material-specific calculations would establish their strength and assess their influence on collective lattice dynamics.

In the lattice considered here, both the unequal correlations generated by symmetric elastic anisotropy and the antisymmetric displacement interaction contribute to countercirculating currents on neighboring plaquettes. These currents are generally present both in the phonon vacuum and at thermal equilibrium. Crystal symmetry fixes their relative pattern, while the elastic couplings determine their magnitude and direction of circulation. The mean angular momentum and total mean crystalline torque vanish at every site, even though anisotropic forces shape the correlations that sustain the currents.

The dependence of the current on intersite displacement correlations provides guidance for its experimental detection. Recent advances in electron ptychography provide spatially resolved access to interatomic displacement correlations~\cite{Gladyshev2026}, suggesting a route to reconstructing the bond currents from the relevant tensor components and independently determined elastic couplings.
 The current is quadratic in
displacement and invariant under the simultaneous reversal of all
displacement and momentum vectors. In harmonic equilibrium, its expectation value consequently has no
linear response to a force coupled linearly to displacement.
Optical changes in vibrational frequencies can generate correlated phonon pairs~\cite{Trigo2013}, while second-order Raman processes provide a route to modifying displacement fluctuations~\cite{PhysRevLett.79.4605,Garrett1997,Henighan2016}.
When a perturbation also changes the exchange coefficients, the
response includes both the altered correlations and the explicit
variation of the current operator. An experimental test thus should
resolve the staggered pattern and distinguish its contribution
from other correlations in the same symmetry channel.

The broader relevance of our findings is sharpened by their
connection to microscopic descriptions of altermagnets and
cuprate $d$-wave superconductors. Rotation-related local crystal
fields shape altermagnetic electronic
structure~\cite{Smejkal2022}, while early oxygen-resolved models of
the cuprates retain the directional character of Cu--O
bonding~\cite{Emery1987,MattheissHamann1989}. Effective theories of oxygen motion further demonstrate how lattice
fluctuations can contribute to interactions in the $d$-wave pairing
channel~\cite{BulutScalapino1996,NewnsTsuei2007}, while recent
first-principles calculations suggest a connection between rotational
oxygen phonons and low-energy electronic spectral anomalies in
cuprates~\cite{Wang2026RotationalPhonons}.  These precedents
identify concrete settings in which to investigate the rotational
correlations of internal lattice degrees of freedom before
electronic or magnetic order develops.

Our work places such correlations within the symmetry-preserving reference state of the coupled many-body problem, suggesting that the crystal supplies more than a background geometry: its own correlated fluctuations may help determine the order that emerges. Theoretical studies have shown how chiral-phonon couplings can mediate electron--electron~\cite{Gao2023} and spin--spin~\cite{Yokoyama2024} interactions. Related proposals for time-reversal-preserving electronic spin-current order~\cite{Raghu2008} further motivate exploring whether lattice circulation can influence electronic ordering. Future studies should determine the relevant dynamical correlations and microscopic couplings to establish whether the rotational structure already present in the fluctuating lattice helps select or stabilize an ordered phase.

\section*{Acknowledgments}
This work was supported by the National Science Foundation under Grant No. NSF DMR-2144086.

\appendix

\section{Static reduction and effective elastic coefficients}
\label{app:parent}
In this Appendix, we express the effective elastic coefficients in
Eq.~\eqref{eq:H} in terms of the spring constants of the parent
model introduced in Sec.~\ref{sec:Model}. We give the parent
matrix blocks, specify the Fourier projections used in the
NN--NNN truncation, and evaluate their leading weak-stiffness
expansions.

The constants $k_{\mathrm{MnMn},L/T}$,
$k_{\mathrm O,L/T}$, $k_{\mathrm{Se},L/T}$, and
$k_{\mathrm{OSe},L/T}$ denote, respectively, the direct
Mn--Mn, Mn--O, Mn--Se, and O--Se spring stiffnesses.
The two Mn masses are $M$, while the internal masses are
$m_{\mathrm O}$ and $m_{\mathrm{Se}}$.
All four spring families and their bond directions are those
specified in Sec.~\ref{sec:Model}. We use the Fourier convention in
Eq.~\eqref{eq:fourier-convention} and define
\begin{equation}
 c_\mu=\cos\frac{k_\mu a}{2},
 \qquad
 s_\mu=\sin\frac{k_\mu a}{2},
 \qquad \mu=x,y,
\label{appA:trigonometric-factors}
\end{equation}
where the Brillouin zone is $[-\pi/a,\pi/a)^2$.

In the Cartesian order
\[
(A_x,A_y,B_x,B_y;\mathrm O_x,\mathrm O_y,
\mathrm{Se}_x,\mathrm{Se}_y),
\]
the parent dynamical matrix can be written as
\begin{equation}
D^{\mathrm{parent}}(\mathbf k)=
\begin{pmatrix}
D_{\mathrm{Mn},\mathrm{Mn}}&D_{\mathrm{Mn},\mathrm{int}}\\
D_{\mathrm{Mn},\mathrm{int}}^{T}&D_{\mathrm{int},\mathrm{int}}
\end{pmatrix}.
\label{appA:parent-block-matrix}
\end{equation}
All blocks below are evaluated at the same $\mathbf k$.
The notation $D_{s,s'}$ in this Appendix denotes a $2\times2$
Cartesian block of the parent matrix, with row and column
order $(x,y)$. The retained block is
\begin{equation}
D_{\mathrm{Mn},\mathrm{Mn}}=
\begin{pmatrix}
D_{A,A}&D_{A,B}\\
D_{A,B}^{T}&D_{B,B}
\end{pmatrix},
\label{appA:parent-Mn-block}
\end{equation}
where
\begin{equation}
\begin{aligned}
D_{A,A}
={}&\frac{2(k_{\mathrm{MnMn},L}+k_{\mathrm{MnMn},T})}{M}
\mathbb I_2
\\
&+\frac{2}{M}
\begin{pmatrix}
k_{\mathrm O,L}+k_{\mathrm{Se},T}&0\\
0&k_{\mathrm O,T}+k_{\mathrm{Se},L}
\end{pmatrix},
\\
D_{B,B}
={}&\frac{2(k_{\mathrm{MnMn},L}+k_{\mathrm{MnMn},T})}{M}
\mathbb I_2
\\
&+\frac{2}{M}
\begin{pmatrix}
k_{\mathrm O,T}+k_{\mathrm{Se},L}&0\\
0&k_{\mathrm O,L}+k_{\mathrm{Se},T}
\end{pmatrix},
\\
D_{A,B}
={}&-\frac{2(k_{\mathrm{MnMn},L}+k_{\mathrm{MnMn},T})}{M}
 c_xc_y\mathbb I_2
\\
&+\frac{2(k_{\mathrm{MnMn},L}-k_{\mathrm{MnMn},T})}{M}
 s_xs_y\begin{pmatrix}0&1\\1&0\end{pmatrix}.
\end{aligned}
\label{appA:parent-Mn-elements}
\end{equation}
The mixed block is
\begin{equation}
D_{\mathrm{Mn},\mathrm{int}}=
\begin{pmatrix}
D_{A,\mathrm O}&D_{A,\mathrm{Se}}\\
D_{B,\mathrm O}&D_{B,\mathrm{Se}}
\end{pmatrix},
\label{appA:parent-mixed-block}
\end{equation}
with
\begin{equation}
\begin{aligned}
D_{A,\mathrm O}
&=-\frac{2c_x}{\sqrt{Mm_{\mathrm O}}}
\begin{pmatrix}k_{\mathrm O,L}&0\\0&k_{\mathrm O,T}\end{pmatrix},
\\
D_{B,\mathrm O}
&=-\frac{2c_y}{\sqrt{Mm_{\mathrm O}}}
\begin{pmatrix}k_{\mathrm O,T}&0\\0&k_{\mathrm O,L}\end{pmatrix},
\\
D_{A,\mathrm{Se}}
&=-\frac{2c_y}{\sqrt{Mm_{\mathrm{Se}}}}
\begin{pmatrix}k_{\mathrm{Se},T}&0\\0&k_{\mathrm{Se},L}\end{pmatrix},
\\
D_{B,\mathrm{Se}}
&=-\frac{2c_x}{\sqrt{Mm_{\mathrm{Se}}}}
\begin{pmatrix}k_{\mathrm{Se},L}&0\\0&k_{\mathrm{Se},T}\end{pmatrix}.
\end{aligned}
\label{appA:parent-mixed-elements}
\end{equation}
Finally,
\begin{equation}
D_{\mathrm{int},\mathrm{int}}=
\begin{pmatrix}
D_{\mathrm O,\mathrm O}&D_{\mathrm O,\mathrm{Se}}\\
D_{\mathrm O,\mathrm{Se}}^{T}&D_{\mathrm{Se},\mathrm{Se}}
\end{pmatrix},
\label{appA:parent-internal-block}
\end{equation}
where
\begin{equation}
\begin{aligned}
D_{\mathrm O,\mathrm O}
&=\frac{2}{m_{\mathrm O}}
\bigl(k_{\mathrm O,L}+k_{\mathrm O,T}
+k_{\mathrm{OSe},L}+k_{\mathrm{OSe},T}\bigr)\mathbb I_2,
\\
D_{\mathrm{Se},\mathrm{Se}}
&=\frac{2}{m_{\mathrm{Se}}}
\bigl(k_{\mathrm{Se},L}+k_{\mathrm{Se},T}
+k_{\mathrm{OSe},L}+k_{\mathrm{OSe},T}\bigr)\mathbb I_2,
\\
D_{\mathrm O,\mathrm{Se}}
&=-\frac{2(k_{\mathrm{OSe},L}+k_{\mathrm{OSe},T})}
{\sqrt{m_{\mathrm O}m_{\mathrm{Se}}}}
 c_xc_y\mathbb I_2
\\
&\quad+\frac{2(k_{\mathrm{OSe},L}-k_{\mathrm{OSe},T})}
{\sqrt{m_{\mathrm O}m_{\mathrm{Se}}}}
 s_xs_y\begin{pmatrix}0&1\\1&0\end{pmatrix}.
 \label{appA:parent-internal-elements}
\end{aligned}
\end{equation}
These blocks completely specify the $8\times8$ parent matrix.

Minimizing the elastic energy with respect to the O and Se
displacements while holding the Mn displacements fixed yields
the statically reduced matrix in Eq.~\eqref{eq:static-reduction}.
The internal mass factors cancel in the product of the mixed
blocks and the inverse internal block, so neither
$D_{\mathrm{stat}}$ nor the effective elastic coefficients
depend on $m_{\mathrm O}$ or $m_{\mathrm{Se}}$.
The Mn mass enters only through an overall factor of $1/M$.
For the NN terms retained in the effective Hamiltonian,
Eq.~\eqref{eq:H} yields
\begin{equation}
[D(\mathbf k)]_{A_x,B_y}-[D(\mathbf k)]_{A_y,B_x}
=8\lambda s_xs_y.
\label{appA:effective-antisymmetric-harmonic}
\end{equation}

Since
$a^2\int_{\mathrm{BZ}}d^2\mathbf k\,
 s_x^2s_y^2/(2\pi)^2=1/4$,
the antisymmetric NN projection of the statically reduced
matrix is

\begin{equation}
\begin{aligned}
\lambda =&
\ \frac{a^2}{2}
\int_{\mathrm{BZ}}\frac{d^2\mathbf{k}}{(2\pi)^2}
\sin\frac{k_xa}{2}\sin\frac{k_ya}{2}
\\
&\times
\left[
D_{\mathrm{stat};A_x,B_y}(\mathbf{k})
-
D_{\mathrm{stat};A_y,B_x}(\mathbf{k})
\right].
\end{aligned}
\label{eq:lambda_projection}
\end{equation}
The symmetric NN projections similarly give
\begin{equation}
\begin{aligned}
\Gamma_1+\Gamma_{1T}
&=-2a^2\int_{\mathrm{BZ}}\frac{d^2\mathbf k}{(2\pi)^2}
 c_xc_y\,D_{\mathrm{stat};A_x,B_x}(\mathbf k),
\\
\Gamma_1-\Gamma_{1T}
&=a^2\int_{\mathrm{BZ}}\frac{d^2\mathbf k}{(2\pi)^2}
 s_xs_y
\\
&\quad\times\bigl[
D_{\mathrm{stat};A_x,B_y}(\mathbf k)
+D_{\mathrm{stat};A_y,B_x}(\mathbf k)
\bigr].
\end{aligned}
\label{appA:symmetric-NN-projections}
\end{equation}
For the axial NNN bonds, we obtain
\begin{equation}
\begin{aligned}
\Gamma_2
&=-a^2\int_{\mathrm{BZ}}\frac{d^2\mathbf k}{(2\pi)^2}
 \cos(k_xa)D_{\mathrm{stat};A_x,A_x}(\mathbf k),
\\
\Gamma_{2T}
&=-a^2\int_{\mathrm{BZ}}\frac{d^2\mathbf k}{(2\pi)^2}
 \cos(k_xa)D_{\mathrm{stat};A_y,A_y}(\mathbf k),
\\
\Gamma^\prime_2
&=-a^2\int_{\mathrm{BZ}}\frac{d^2\mathbf k}{(2\pi)^2}
 \cos(k_ya)D_{\mathrm{stat};A_y,A_y}(\mathbf k),
\\
\Gamma^\prime_{2T}
&=-a^2\int_{\mathrm{BZ}}\frac{d^2\mathbf k}{(2\pi)^2}
 \cos(k_ya)D_{\mathrm{stat};A_x,A_x}(\mathbf k).
\end{aligned}
\label{appA:axial-NNN-projections}
\end{equation}
The corresponding $B$-sublattice projections follow from
Eq.~\eqref{appA:axial-NNN-projections} by the fourfold rotation
that interchanges the two Mn sublattices.
Together, Eqs.~\eqref{eq:lambda_projection}--\eqref{appA:axial-NNN-projections}
extract from $D_{\mathrm{stat}}$ the NN and axial NNN couplings
retained in the effective Hamiltonian~\eqref{eq:H}.

To obtain explicit expressions for these couplings, we expand
the static reduction~\eqref{eq:static-reduction} in
$k_{\mathrm O,T}$, $k_{\mathrm{Se},T}$,
$k_{\mathrm{OSe},L/T}$, and $k_{\mathrm{MnMn},L/T}$,
while holding $k_{\mathrm O,L}$ and $k_{\mathrm{Se},L}$
positive and fixed. Each expansion parameter is assumed small
compared with both longitudinal Mn--ligand stiffnesses,
and all are assigned the same perturbative order.
Substituting the expanded matrix into
Eqs.~\eqref{appA:symmetric-NN-projections}
and~\eqref{appA:axial-NNN-projections} yields, through first order,
the NN and longitudinal axial NNN coefficients
\begin{equation}
\begin{aligned}
\Gamma_1 \simeq{}&
\frac{k_{\mathrm{MnMn},L}}{M}
+
\frac{k_{\mathrm{O},T}+k_{\mathrm{Se},T}}{2M}
+
\frac{3(k_{\mathrm{OSe},L}+k_{\mathrm{OSe},T})}{8M},
\\[3pt]
\Gamma_{1T} \simeq{}&
\frac{k_{\mathrm{MnMn},T}}{M}
+
\frac{k_{\mathrm{O},T}+k_{\mathrm{Se},T}}{2M}
+
\frac{3(k_{\mathrm{OSe},L}+k_{\mathrm{OSe},T})}{8M},
\\[3pt]
\Gamma_2 \simeq{}&
\frac{
k_{\mathrm{O},L}-k_{\mathrm{O},T}
-k_{\mathrm{OSe},L}-k_{\mathrm{OSe},T}
}{2M},
\\[3pt]
\Gamma^\prime_2 \simeq{}&
\frac{
k_{\mathrm{Se},L}-k_{\mathrm{Se},T}
-k_{\mathrm{OSe},L}-k_{\mathrm{OSe},T}
}{2M}.
\end{aligned}
\label{eq:effective_symmetric_couplings}
\end{equation}
Evaluating the transverse projections in
Eq.~\eqref{appA:axial-NNN-projections} to second order yields
the leading nonzero contributions to $\Gamma_{2T}$ and
$\Gamma^\prime_{2T}$, i.e.,
\begin{equation}
\begin{aligned}
\Gamma_{2T} \simeq{}&
\frac{
(2k_{\mathrm{O},T}+k_{\mathrm{OSe},L}
+k_{\mathrm{OSe},T})^2
+2k_{\mathrm{OSe},L}k_{\mathrm{OSe},T}
}{
8Mk_{\mathrm{O},L}
},
\\[5pt]
\Gamma^\prime_{2T} \simeq{}&
\frac{
(2k_{\mathrm{Se},T}+k_{\mathrm{OSe},L}
+k_{\mathrm{OSe},T})^2
+2k_{\mathrm{OSe},L}k_{\mathrm{OSe},T}
}{
8Mk_{\mathrm{Se},L}
}.
\end{aligned}
\label{eq:effective_transverse_couplings}
\end{equation}
At the same order, Eq.~\eqref{eq:lambda_projection} yields
the leading contribution to the antisymmetric NN coupling:
\begin{equation}
\begin{aligned}
\lambda \simeq&
\frac{
k_{\mathrm{OSe},L}-k_{\mathrm{OSe},T}
}{
32Mk_{\mathrm{O},L}k_{\mathrm{Se},L}
}
\Bigl[
(k_{\mathrm{OSe},L}+k_{\mathrm{OSe},T})
\\
&\times
(k_{\mathrm{O},L}-k_{\mathrm{Se},L})
+
4\bigl(
k_{\mathrm{O},L}k_{\mathrm{Se},T}
-k_{\mathrm{O},T}k_{\mathrm{Se},L}
\bigr)
\Bigr].
\end{aligned}
\label{eq:effective_antisymmetric_coupling}
\end{equation}

For $k_{\mathrm{O},T}=k_{\mathrm{Se},T}=0$,
Eq.~\eqref{eq:effective_antisymmetric_coupling} reduces to
Eq.~\eqref{eq:lambda}. Equal but nonzero transverse Mn--ligand
stiffnesses do not generally eliminate the antisymmetric coupling
if the longitudinal stiffnesses remain unequal.
By contrast, imposing both
$k_{\mathrm{O},L}=k_{\mathrm{Se},L}$ and
$k_{\mathrm{O},T}=k_{\mathrm{Se},T}$ restores the equivalence
of the two elastic environments, forcing $\lambda=0$ by symmetry.
The coupling also vanishes when the longitudinal and transverse
stiffnesses are equal on every bond of the parent network;
this cancellation is exact and does not rely on the
perturbative expansion.

\section{Dynamical matrix and current expectation value}
\label{app:DynamicalMatrix}
In this Appendix, we derive the dynamical matrix entering
Eq.~\eqref{PhononEigenproblem} and establish an exact sign rule
for the equilibrium current in Eq.~\eqref{curre} at $\lambda=0$.
The sign rule extends the perturbative result in
Eq.~\eqref{eq:expansion} beyond small stiffness contrasts
and to finite temperature.
Using the Fourier convention in Eq.~\eqref{eq:fourier-convention}
and the trigonometric factors defined in
Eq.~\eqref{appA:trigonometric-factors}, we obtain from the
Hamiltonian in Eq.~\eqref{eq:H} the following dynamical matrix
in the Cartesian basis $(A_x,A_y,B_x,B_y)$:
\begin{equation}
\hspace{-1.2em} 
\begin{aligned}
D(\mathbf k)
&=\begin{pmatrix}
D_A(\mathbf k) & D_{AB}(\mathbf k)\\
D_{AB}^{T}(\mathbf k) & D_B(\mathbf k)
\end{pmatrix},
\\
D_A(\mathbf k)
&=2(\Gamma_1+\Gamma_{1T})\mathbb I_2
+4s_x^2\begin{pmatrix}\Gamma_2&0\\0&\Gamma_{2T}\end{pmatrix}
+4s_y^2\begin{pmatrix}\Gamma^\prime_{2T}&0\\0&\Gamma^\prime_2\end{pmatrix},
\\
D_B(\mathbf k)
&=2(\Gamma_1+\Gamma_{1T})\mathbb I_2
+4s_x^2\begin{pmatrix}\Gamma^\prime_2&0\\0&\Gamma^\prime_{2T}\end{pmatrix}
+4s_y^2\begin{pmatrix}\Gamma_{2T}&0\\0&\Gamma_2\end{pmatrix},
\\
D_{AB}(\mathbf k)
&=-2(\Gamma_1+\Gamma_{1T})c_xc_y\mathbb I_2
\\
&\quad+2s_xs_y\begin{pmatrix}
0&(\Gamma_1-\Gamma_{1T})+2\lambda\\
(\Gamma_1-\Gamma_{1T})-2\lambda&0
\end{pmatrix}.
\end{aligned}
\label{appB:dynamical-matrix}
\end{equation}
Translational invariance enforces two zero-frequency acoustic modes at
$\mathbf{k}=0$. We assume mechanical stability, with strictly positive
squared frequencies for all modes other than these two uniform translations.

For the remainder of this Appendix, we set $\lambda=0$ and parametrize
the transverse NNN couplings by a common ratio $\rho$,
$\Gamma_{2T}=\rho\Gamma_2$ and
$\Gamma^\prime_{2T}=\rho\Gamma^\prime_2$,
with $0\leq\rho<1$ and
$\Gamma_1,\Gamma_{1T},\Gamma_2,\Gamma^\prime_2>0$.
The thermal weight in Eq.~\eqref{eq:thermal-weight} admits the
resolvent representations
\begin{equation}
\begin{aligned}
\frac{\hbar}{2}g_0(\omega)
&=\frac{\hbar}{\pi}
\int_0^\infty\frac{d\xi}{\omega^2+\xi^2},
\\
\frac{\hbar}{2}g_T(\omega)
&=k_BT\sum_{m=-\infty}^{\infty}
\frac{1}{\omega^2+\xi_m^2},
\\
\xi_m&=\frac{2\pi m k_BT}{\hbar},
\end{aligned}
\label{appB:weight-resolvent}
\end{equation}
where $\omega>0$, and the second identity applies for $T>0$.
Inserting these identities into Eq.~\eqref{curre} and using
Eq.~\eqref{PhononEigenproblem} to sum the polarization projectors
gives 
\begin{equation}
\begin{aligned}
I(0)={}&-\frac{\hbar (\Gamma_1+\Gamma_{1T})}{2\pi}
\int_0^\infty d\xi
\int_{\mathrm{BZ}}\frac{a^2d^2k}{(2\pi)^2}
\\
&\times\cos(\mathbf k\cdot\boldsymbol\delta_{++})
\Bigl\{[D(\mathbf k)+\xi^2\mathbb I_4]^{-1}_{Ax,By}
\\
&\qquad-[D(\mathbf k)+\xi^2\mathbb I_4]^{-1}_{Ay,Bx}\Bigr\}.
\end{aligned}
\label{appB:current-resolvent}
\end{equation}

The crossed inverse elements in Eq.~\eqref{appB:current-resolvent} obey the exact  identity
\begin{equation}
\begin{aligned}
&[D(\mathbf k)+\xi^2\mathbb I_4]^{-1}_{Ax,By}
-[D(\mathbf k)+\xi^2\mathbb I_4]^{-1}_{Ay,Bx}
\\
&\quad=\frac{2(\Gamma_1-\Gamma_{1T})s_xs_y}
{\det[D(\mathbf k)+\xi^2\mathbb I_4]}
\\
&\qquad\times\Bigl\{
[D_{Ax,Ax}(\mathbf k)+\xi^2]
[D_{By,By}(\mathbf k)+\xi^2]
\\
&\quad\qquad-
[D_{Ay,Ay}(\mathbf k)+\xi^2]
[D_{Bx,Bx}(\mathbf k)+\xi^2]
\Bigr\}.
\end{aligned}
\label{appB:cofactor}
\end{equation}
Substituting the diagonal entries of the dynamical matrix $D(\mathbf{k})$
from Eq.~\eqref{appB:dynamical-matrix} yields
\begin{equation}
\begin{aligned}
&[D_{Ax,Ax}(\mathbf k)+\xi^2]
[D_{By,By}(\mathbf k)+\xi^2]
\\
&\quad-
[D_{Ay,Ay}(\mathbf k)+\xi^2]
[D_{Bx,Bx}(\mathbf k)+\xi^2]
\\
&\quad=4(1-\rho)(\Gamma_2-\Gamma^\prime_2)
\\
&\qquad\times\Bigl\{
[2(\Gamma_1+\Gamma_{1T})+\xi^2](s_x^2+s_y^2)
\\
&\qquad\qquad+
4(1+\rho)(\Gamma_2+\Gamma^\prime_2)s_x^2s_y^2
\Bigr\}.
\end{aligned}
\label{appB:factorization}
\end{equation}
Under the stated assumptions, the expression in braces in
Eq.~\eqref{appB:factorization} is strictly positive whenever
$s_x^2+s_y^2>0$ and is even in each momentum component.
For the reference bond, the Fourier factor is
$\cos(\mathbf k\cdot\boldsymbol\delta_{++})=c_xc_y-s_xs_y$.
In the Cartesian basis $(A_x,A_y,B_x,B_y)$, the matrix
$S=\operatorname{diag}(1,-1,1,-1)$ satisfies
\begin{equation}
S D(k_x,k_y) S = D(-k_x,k_y) = D(k_x,-k_y).
\end{equation}
Thus, $\det[D(\mathbf k)+\xi^2\mathbb I_4]$ is even in each
momentum component. The term proportional to $c_xc_ys_xs_y$
in the current integrand is odd under either reflection and
integrates to zero over the Brillouin zone. Only the $-s_xs_y$
part of the Fourier factor therefore contributes to the current.

Combining Eqs.~\eqref{appB:current-resolvent}--\eqref{appB:factorization}
yields
\begin{equation}
\begin{aligned}
I(0)={}&
\frac{4\hbar (\Gamma_1+\Gamma_{1T})(\Gamma_1-\Gamma_{1T})(1-\rho)(\Gamma_2-\Gamma^\prime_2)}{\pi}
\\
&\times\int_0^\infty d\xi
\int_{\mathrm{BZ}}\frac{a^2d^2k}{(2\pi)^2}
\frac{s_x^2s_y^2}{\det[D(\mathbf k)+\xi^2\mathbb I_4]}
\\
&\times\Bigl\{
[2(\Gamma_1+\Gamma_{1T})+\xi^2](s_x^2+s_y^2)
\\
&\qquad+
4(1+\rho)(\Gamma_2+\Gamma^\prime_2)s_x^2s_y^2
\Bigr\}.
\end{aligned}
\label{appB:vacuum-current}
\end{equation}
Mechanical stability ensures that the denominator is strictly positive
for $\mathbf k\ne0$ and $\xi\geq0$. The integrand is positive wherever
$s_xs_y\ne0$, so the integral is strictly positive.
At finite temperature, replacing $(\hbar/\pi)\int_0^\infty d\xi$
by $k_BT\sum_{m=-\infty}^{\infty}$, with the integrand evaluated at
$\xi=\xi_m$, preserves this positivity.
Consequently, for all $T\geq0$,
\begin{equation}
\left.\operatorname{sgn}I(T)\right|_{\lambda=0}
=\operatorname{sgn}\bigl[
(\Gamma_1-\Gamma_{1T})(\Gamma_2-\Gamma^\prime_2)
\bigr].
\label{eq:sign}
\end{equation}
This sign rule is exact under the stated assumptions and requires
neither a small-contrast nor a long-wavelength approximation.



\section{Local rotational symmetry and an upper bound on bond currents}
\label{app:bond-current-bound}

In this Appendix, we derive the upper bound in
Eq.~\eqref{eq:bond-current-upper-bound} on the vacuum expectation
value of the NN current defined in Eq.~\eqref{source:current}.
The derivation uses the invariant bond energy in Eq.~\eqref{Hinv}
and the angular momenta at the bond endpoints, introduced in
Sec.~\ref{sec:AM}.
Combining Eq.~\eqref{source:current} with the common-rotation
identity in Eq.~\eqref{365eq} yields
\begin{equation}
\begin{aligned}
\bigl[\hat h_{AB}^{\mathrm{inv}},\hat L_A\bigr]
&= i\hbar\hat J_{A\to B},\\
\bigl[\hat h_{AB}^{\mathrm{inv}},\hat L_B\bigr]
&=-i\hbar\hat J_{A\to B}.
\end{aligned}
\label{appC:endpoint-commutators}
\end{equation}
Define the bond-energy fluctuation
\[
\delta\hat h_{AB}^{\mathrm{inv}}
\equiv \hat h_{AB}^{\mathrm{inv}}
-\langle\hat h_{AB}^{\mathrm{inv}}\rangle_0.
\]
For either endpoint $s=A,B$, Hermiticity implies
\begin{equation}
\langle\hat L_s\,\delta\hat h_{AB}^{\mathrm{inv}}\rangle_0
=\langle\delta\hat h_{AB}^{\mathrm{inv}}\,\hat L_s\rangle_0^{*}.
\label{appC:conjugation}
\end{equation}
Thus Eq.~\eqref{appC:endpoint-commutators} implies
\begin{equation}
 \frac{\hbar}{2}
 \bigl|\langle\hat J_{A\to B}\rangle_0\bigr|
 =\bigl|\operatorname{Im}
 \langle\delta\hat h_{AB}^{\rm inv}\,\hat L_s\rangle_0\bigr|.
 \label{appC:current-imaginary-part}
\end{equation}
Applying the Cauchy--Schwarz inequality to
$\delta\hat h_{AB}^{\rm inv}|0\rangle$ and $\hat L_s|0\rangle$
yields
\begin{equation}
 \bigl|\langle\delta\hat h_{AB}^{\rm inv}\,\hat L_s\rangle_0\bigr|^2
 \le
 \langle(\delta\hat h_{AB}^{\rm inv})^2\rangle_0
 \langle\hat L_s^2\rangle_0.
 \label{appC:cauchy-schwarz}
\end{equation}
The bound in Eq.~\eqref{eq:bond-current-upper-bound} for $i=A$ and
$j=B$ follows from Eqs.~\eqref{appC:current-imaginary-part} and
\eqref{appC:cauchy-schwarz} by bounding the magnitude of the
imaginary part by the modulus.

This inequality is the Robertson uncertainty bound applied to the
invariant bond energy and an endpoint angular momentum~\cite{Robertson1929},
with $\langle\hat L_s\rangle_0=0$.
If the bond-energy second moment is finite, the bound implies that
a nonzero mean current requires nonzero angular momentum fluctuations
at both endpoints.
The staggered angular momentum
$\hat L_{\mathrm{st}}=\hat L_A-\hat L_B$, introduced in
Sec.~\ref{sec:Current}, obeys
\[
\bigl[\hat h_{AB}^{\mathrm{inv}},\hat L_{\mathrm{st}}\bigr]
=2i\hbar\hat J_{A\to B}.
\]
Applying the same argument to $\hat L_{\mathrm{st}}$ yields
the relative-rotation form of the bound,
\begin{equation}
\begin{aligned}
 \bigl|\langle\hat J_{A\to B}\rangle_0\bigr|
 \le \frac{1}{\hbar}
 \Bigl[
 &\bigl(\langle(\hat h_{AB}^{\rm inv})^2\rangle_0
       -\langle\hat h_{AB}^{\rm inv}\rangle_0^2\bigr)
 \\
 &\times\langle(\hat L_A-\hat L_B)^2\rangle_0
 \Bigr]^{1/2}.
\end{aligned}
\label{appC:relative-current-upper-bound}
\end{equation}
The different prefactor in this bound reflects the factor of two
in the commutator.

The relevant symmetry is invariance of the complete Hamiltonian
under independent rotations of a single site, with all other sites,
bond directions, and background sources held fixed.
If $[\hat H,\hat L_s]=0$, a nondegenerate ground state must also
be an eigenstate of $\hat L_s$. Time-reversal symmetry then forces
the corresponding eigenvalue to vanish, so the bound in
Eq.~\eqref{eq:bond-current-upper-bound} vanishes. Hence
\begin{equation}
\begin{aligned}
 [\hat H,\hat L_s]=0
 &\ \Longrightarrow\ \hat L_s|0\rangle=0
 \\
 &\ \Longrightarrow\ \langle\hat J_{A\to B}\rangle_0=0,
 \quad s=A\ \text{or}\ B.
\end{aligned}
 \label{appC:local-no-current}
\end{equation}
Conservation of $\hat L_{\mathrm{st}}$ leads to the same conclusion
through Eq.~\eqref{appC:relative-current-upper-bound}.
These conclusions rely on local or relative rotational symmetries.
Equation~\eqref{365eq}, in contrast, expresses invariance under
a common rotation of the two endpoints while allowing angular
momentum exchange between them. Likewise, setting $\hat H^\tau=0$
restores the global $U(1)$ symmetry generated by $\hat L$ without
restoring independent rotational symmetries at the endpoints.
In this limit, Eq.~\eqref{eq:balance} requires the stationary
current to be divergence-free but permits nonzero current
expectation values on individual bonds.
The current induced by $\lambda$ realizes this possibility.

\bibliographystyle{apsrev4-2}
\bibliography{loops}
\end{document}